\documentclass[11pt]{article}
\usepackage{epsfig}

\newcommand{\be}{\begin{eqnarray}}
\newcommand{\ee}{\end{eqnarray}}

\def\fr{\frac{1}{2}}

\def\bd{\begin{displaymath}}
\def\ed{\end{displaymath}}
\def\k{\mbox{$\kappa$}}
\def\bo{\begin{array}{c}}
\def\ba12{\begin{array}{cc}}
\def\ea{\end{array}}
\def\nn{\nonumber}
\newfont{\Bbb}{msbm10 scaled 1200}

\begin{document}

\begin{center}

{\LARGE\bf Adiabatic limit interference effects for\\[0.1cm]
two energy level transition amplitudes\\[0.1cm]
and Nikitin - Umanskii formula studied\\[0.5cm]
by fundamental solution method}
\vskip 100pt

{\large {\bf Stefan Giller}}

\vskip 10pt

Institute of Physics, Pedagogical University of Czestochowa\\
ul. Armii Krajowej 13/15; 42-200 Czestochowa, Poland\\
e-mail: sgiller@krysia.uni.lodz.pl
\end{center}
\vspace{60pt}
\begin{abstract}A method of fundamental solutions has been used to
  study adiabatic transition amplitudes in two energy level systems
  for a class of Hamiltonians allowing some simplifications of Stokes graphs
  corresponding to such transitions. It has been shown that for simplest such
  cases the amplitudes take the Nikitin - Umanskii form but for more complicated ones
  they are formed by a sum of terms strictly related to a structure of Stokes graph
  corresponding to such cases. This paper corrects our previous one [Phys. Rev. A,
  {\bf 63} 052101 (2001)] and its results are in a full agreement with the ones of
   Joye, Mileti and Pfister [Phys. Rev. A, {\bf 44} 4280 (1991)].
\end{abstract}
\vskip 9pt

{\small PACS number(s): 03.65.-W , 03.65.Sq , 02.30.Lt , 02.30.Mv}


\vskip 20pt

\section{Introduction}

\hskip+2em In our previous paper \cite{01} we have applied a formalism of fundamental
solutions to obtain formulae for adiabatic transition amplitudes in two level
energy systems. The corresponding formalism has been developed under quite general
assumptions about a nature of  Hamiltonians perturbing a system adiabatically.
Unfortunately, in its applications to particular examples considered in the section V
and the furthers of the paper mentioned we have made a fatal error in detailed
calculations of the corresponding transition
amplitudes.
As a consequence of this we have also drawn in this paper erroneous conclusions which
followed from the obtained erroneous formulae.
In the present paper we would like to correct the corresponding
calculations as well as to draw correct conclusions.

However in order to avoid a permanent referring to the material presented and
discussed in the first
four sections of the paper \cite{01} and to make our present paper selfsufficient
and selfconsistent we shall repeat below to large extent the contents of these
sections.
Therefore we shall start with reminding shortly main reasons for studying
transitions in two energy level systems.

First of all such systems provide us with the simplest models
to investigate transition amplitudes between
different energy levels by different approaches \cite{1}. On the other side these systems play an important role in
experimental investigations of basic principles of quantum
 mechanics \cite{2}. Recently a lot of effort has been devoted to obtain more rigorous results on
the adiabatic limit of transition amplitudes for these systems
 \cite{3,4,5,6,7}. In particular in a series
of recent papers Joye {\it et al} \cite{3,4,5,6,7} have studied this problem by the Hilbert
space methods. Such
two energy level systems are formally equivalent to a one-half spin system put into
time
dependent magnetic field. However  good approximate results and the more so the exact
 ones are difficult to
obtain for such systems even for simple time evolutions of the effective 'magnetic'
field. Therefore each opportunity of improving this situation is worth trying.
A treatment of the problem by a method of fundamental solutions (so fruitful in its
application
to stationary problems of 1-dim Schr\"odinger equation \cite{8,9,10}) is of first
importance, the more
so that to our knowledge, the method was not used so far to
 this goal. A possibility
of application of the method is related to the fact that a linear system of first
order
differential equations describing time evolution of transition
 amplitudes can always be transformed into a system of decoupled second order
equations having a form of the stationary Schr\"odinger equation, one for each
amplitude. This allows us to apply all advantages of the fundamental solution
method \cite{10}. The
only obstacle related with this approach is a complexity of effective 'potentials'
which appear
in the final system of the Schr\"odinger-type equations.

     The paper is organized as follows.

     In the next section the problem of transitions in two energy level systems is stated
and corresponding assumptions about the effective 'magnetic field' are formulated. A linear
system of two differential equations for the transition amplitudes is rewritten
in a form of two decoupled equations of the Schr\"odinger type.

     In Sec.3 properties of the fundamental solution method are recalled.

     In Sec.4 some subtleties of the application of the fundamental solution method to the
problems considered in the paper are discussed.

In Sec.5 a class of Hamiltonians with so called NED property is distinguished for
further considerations.

In Sec.6 an exact form of a transition amplitude for the NED systems is obtained
and its adiabatic limit is found.

In Sec.7 two examples of the NED systems are considered and the Nikitin - Umanskii
formula is reconstructed.

 Finally in Sec.8 we discuss our results stressing their coincidence with the
 corresponding ones of Joye {\it et al} \cite{4}.

\section{Adiabatic transitions in two energy level systems}

\hskip+2em First let us remind that, in general,
any two energy level system is formally equivalent to a
one-half spin system put into an external magnetic field ${\bf
  B}(t)$. Its
Hamiltonian $H(t)$ is given then by $H(t)=\frac{1}{2}\mu {\bf
  B}(t)\cdot{\bf \sigma}$ , where ${\bf \sigma}=(\sigma_x,\sigma_y,\sigma_z)$ are Pauli's matrices so
that two energy levels $E_{\pm}(t)$ of $H(t)$ are given by
$E_{\pm}(t)=\pm \frac{\mu}{2}B(t)$ where $B(t)=\sqrt{{\bf B}^2(t)}$.

When adiabatic transitions between the two energy levels $E_{\pm}(t)$
 are considered then
the following properties of the field ${\bf B}(t)$ are typically
 assumed:

	${\bf 1}^0$ ${\bf B}(t)$ is real being defined for the real $t$, $-\infty  <t<+\infty$;

	${\bf 2}^0$ ${\bf B}(t)$ can be continued analytically off the real values of $t$
as a meromorphic function
defined on some $t$-Riemann surface ${\bf R}_B$. A sheet of ${\bf R}_B$
 from which ${\bf B}(t)$ is originally     continued is called physical;

	${\bf 3}^0$ On the physical sheet ${\bf B}(t)$ is analytic in an infinite strip
 $\Sigma=\{t:\vert \Im t \vert<\delta, \delta>0\}$,
 without roots in the strip and achieves there finite limits for
$\Re t=\pm \infty$ , i.e.  ${\bf  B}(\Re t=\pm \infty)=
{\bf B}_\pm\ne{\bf 0}$ in the strip;

       ${\bf 4}^0$ $B(t)=\sqrt{{\bf B}^2(t)}$ is a ramified function of $t$ on
       ${\bf R}_B$ with square root branch points coinciding with crossing points
        of the two energy levels $E_{\pm}(t)$;

    The field ${\bf B}(t)$ depends additionally on a parameter $T (>0)$
 i.e. ${\bf B}(t)\equiv{\bf B}(t,T)$ which
introduces a "natural" scale of time to the system, so that its time
 evolution is
expressed most naturally in units of $T$. If $T$ is small in comparison with
the actual period of the
process considered then the latter is "fast" or "sudden". If,
however, $T$ is large in
this comparison then the process is "slow" or "adiabatic".

In the adiabatic process of the system the following is
  assumed about ${\bf B}(t,T)$:

	${\bf 5}^0$ A dependence of ${\bf B}(t,T)$ on $T$ is such that a rescaled
field ${\bf B}(sT,T)$ has the
     following asymptotic behavior for $T\to +\infty$
\begin{eqnarray}
{\bf B}(sT,T)\sim{\bf B}_0(s)+\frac{1}{T}{\bf B}_1(s)+
\frac{1}{T^2}{\bf B}_2(s)+\cdots
\label{2.1}
\end{eqnarray}
    while its $s$-Riemann surface ${\bf R}_B(T)$ approaches 'smoothly' the topological
     structure of the Riemann surface  corresponding to the first term
${\bf B}_0(s)$ of the     expansion (\ref{2.1}).

	${\bf 6}^0$ With respect to its dependence
on $s$ the field ${\bf B}_0(s)$ satisfies properties ${\bf 1}^0-{\bf 4}^0$ above
     with substitutions $t\to s$ and ${\bf B}(s)\to{\bf B}_0(s)$.

	${\bf 7}^0$ For purposes of this paper we shall assume also an algebraic
	dependence of
${\bf B}(sT,T)$ on $s$ so that its asymptotic behaviour in the strip $\Sigma$ as
$s\to\pm\infty$ is the following:
\begin{eqnarray}
{\bf B}(sT,T)\sim {\bf B}_{\pm}+\frac{{\bf
    B}_1^{\pm}}{s^{\alpha_1}}+\frac{{\bf
      B}_2^{\pm}}{s^{\alpha_2}}+\ldots+\frac{{\bf
        B}_k^{\pm}}{s^{\alpha_k}}+\ldots\nn\\
\\
\frac{1}{2}<\alpha_1<\alpha_2<\ldots<\alpha_k<\ldots\hspace{50mm}\nn
\label{6.10}
\end{eqnarray}
where $\alpha_1,\ldots,\alpha_k$, are assumed to be rational.

The next assumption which validity becomes clear in Sec.6 needs to formulate a notion
of Stokes lines for the function $B_0(s)$. These are lines which starts at roots of
$B_0(s)$ and are governed by conditions $\Re\left(i\int_{s_k}^sB_0(\sigma)d\sigma
\right)=0$ where $s_k,\;k=1,...$, are roots of $B_0(s)$. We shall assume the
following about roots of $B_0(s)$, its Stokes lines and about components of the
limiting field ${\bf }B_0(s)$.

${\bf 8}^0$

{bf a}) Roots of $B^2(sT,T)$ and $B_0^2(s)$ are simple.

{\bf b}) A boundary of the central strip
$\Sigma=\{t:\vert \Im t \vert<\delta, \delta>0\}$ in which $B(sT,T)$ is holomorphic
and with no roots inside it consists of two infinite Stokes lines which become
complex conjugated with each other in the adiabatic limit.

{\bf c}) Each component of ${\bf B}(sT,T)$ and ${\bf B}_0(s)$ is holomorphic and
nonvanishing at roots of $B^2(sT,T)$ and $B_0^2(s)$ respectively.

     The time-dependent Schr\"odinger equation corresponding to $H(t)$ takes
a form
\begin{eqnarray}
\frac{i}{T}\frac{d\Psi(s,T)}{ds}=\frac{1}{2} \mu{\bf B}(sT,T)\cdot{\bf \sigma}
\Psi(s,T)
\label{2.2}
\end{eqnarray}

    The adiabatic regime of evolution of the wave function $\Psi(s,T)$
 corresponds now to taking a limit $T\to +\infty$ in (\ref{2.2}).

     The main problem of the adiabatic limit in the considered case is to
find in this limit the
transition amplitude between the two energy levels of the system for
$s\to +\infty$ under the
assumptions that $\Psi(-\infty,T)$ coincides with one of the two possible
eigenstates $\Psi_{\pm}(-\infty,T)$ of $H(-\infty)$ ($=H(+\infty)$) (corresponding to
$E_\pm(-\infty)$ ($=E_\pm(+\infty)$)
and that there is no level crossing for real $t$ i.e.
$\displaystyle\liminf_{-\infty <t <+\infty}B(t)\ge\epsilon>0$.
Known approximate solutions of this
problem are that of Landau \cite{13} and Zener \cite{14} in a form of
the so
called Landau-Zener formula and that of Dykhne \cite{15} who have shown that
such an amplitude should be exponentially small in the limit $T\to +\infty$.
 In the next sections we shall show how
to get an exact (i.e. not approximate) result for this
amplitude as well as its adiabatic limit with the help of the fundamental solutions .

A typical way of proceeding when the adiabatic limit is investigated is
using eigenvectors $\Psi_{\pm}(s,T)$ of $H(sT,T)$ satisfying
$(\Psi_{\pm},\dot\Psi_{\pm})=0$. Then, such eigenvectors
$\Psi_{\pm}(s,T)$ can be chosen as the following ones
\begin{eqnarray}
\Psi_+(s,T)=e^{-i\int
_0^s\dot\phi\sin^2\frac{\Theta}{2}d\sigma}
\left[\begin{array}{c}\cos\frac{\Theta}{2}\\
\\\sin\frac{\Theta}{2}e^{i\phi}\end{array}\right],\hspace{5mm}
\Psi_-(s,T)=e^{-i\int
_0^s\dot\phi\cos^2\frac{\Theta}{2}d\sigma}
\left[\begin{array}{c}\sin\frac{\Theta}{2}\\
\\-\cos\frac{\Theta}{2}e^{i\phi}\end{array}\right]
\label{2.3}
\end{eqnarray}
where $\Theta$ and $\phi$ are respective polar and azimuthal angles
of the vector ${\bf B}(t,T)$ and dots
over different quantities mean derivatives with respect to $s$-variable.

     The wave function $\Psi(s,T)$ can now be represented as
\begin{eqnarray}
\Psi(s,T)=a_+(s,T)e^{-iT \int_0^sE_+(\xi,T)d\xi}\Psi_+(s,T)
+a_-(s,T)e^{-iT \int_0^sE_-(\xi,T)d\xi}\Psi_-(s,T)
\label{2.4}
\end{eqnarray}

The Schr\"odinger equation (\ref{2.2}) can be rewritten in terms of the coefficients $a_{\pm}(s,T)$ as the
following linear system of two equations
\begin{eqnarray}
\dot a_+(s,T)=-c^*(s,T)e^{i\int_0^s\omega(\xi,T)d\xi}a_-(s,T)\nn  \\
\label{2.5}\\
\dot a_-(s,T)=c(s,T)e^{-i\int_0^s\omega(\xi,T)d\xi}a_+(s,T)\nn
\end{eqnarray}
where
\begin{eqnarray}
c(s,T)=  \frac{\dot\Theta}{2} + \frac{i\dot\phi}{2}
\sin\Theta=-\frac{1}{2}\frac{\left[{\bf B}\times\left({\bf
      B}\times{\bf {\dot
        B}}\right)\right]_z}{B^2\sqrt{B_x^2+B_y^2}}+\frac{i}{2}\frac{\left({\bf B}\times{\bf{\dot B}}\right)_z}{ B\sqrt{B_x^2+B_y^2}}\nn\\
\label{2.6}\\
\omega(s,T)=T\left( E_+-E_- \right)-\dot\phi\cos\Theta=\mu
TB-\frac{B_z}{B}\frac{\left({\bf B}\times{\bf{\dot
        B}}\right)_z}{B_x^2+B_y^2}\nn
\end{eqnarray}

According to our assumptions we are looking for a solution to the system (\ref{2.5})
satisfying the following initial conditions $a_+(-\infty,T)=1$ and $\;a_-(-\infty,T)=0$
and under this condition we are interested in the limits
$\lim_{s\to +\infty}a_-(s,T)$ and $\lim_{T\to +\infty}a_-(+\infty,T)$.

It is seen from (\ref{2.6}) however that a dependence of the coefficients $c$ and
$\omega$ on $s$ and $T$ can be quite complicated since the corresponding dependence
of these coefficients on the {\bf B}-field components is sufficiently complicated. We
can simplify the latter dependence by a suitable unitary transformation applied to
$\Psi(s,T)$ in the formula (\ref{2.4}) defined by the following operator
\be
U=e^{\fr i\mu T\int_0^sB_z(sT,T)ds\sigma_z}=
\left[\ba12
e^{\fr i\mu T\int_0^sB_z(sT,T)ds}&0\\&\\0&e^{-\fr i\mu T\int_0^sB_z(sT,T)ds}\ea\right]
\label{2.60}
\ee

For the new amplitudes $\left[\bo a_1(s,T)\\a_2(s,T)\ea\right]$ we get
\be
a_1(s,T)=e^{-i\int_0^s(\dot\phi+\mu TB)\sin^2\frac{\Theta}{2}d\sigma}
\cos\frac{\Theta}{2}a_+(s,T)+\nn
\\e^{-i\int_0^s(\dot\phi-\mu TB)\cos^2\frac{\Theta}{2}d\sigma}
\sin\frac{\Theta}{2}a_-(s,T)\nn
\label{2.62}
\\
\\a_2(s,T)=e^{i\int_0^s(\dot\phi-\mu TB)\cos^2\frac{\Theta}{2}d\sigma+
i\phi(0)}\sin\frac{\Theta}{2}a_+(s,T)-\nn
\\e^{i\int_0^s(\dot\phi+\mu TB)\sin^2\frac{\Theta}{2}d\sigma+
i\phi(0)}\cos\frac{\Theta}{2}a_-(s,T)\nn
\ee

The transformation (\ref{2.60}) does not change the form of Eq.(\ref{2.5})
changing only the corresponding functions $c$ and $\omega$. Namely we have
\be
\dot a_1(s,T)=-Tc_1^*(s,T)e^{iT\int_0^s\omega_1(\xi,T)d\xi}a_2(s,T)\nn
\\\nn
\\\dot a_2(s,T)=Tc_1(s,T)e^{-i\int_0^s\omega_1(\xi,T)d\xi}a_1(s,T)
\label{2.63}
\ee
where
\be
c_1=-\frac{i}{2}\mu B
\sin\Theta e^{i\phi},\;\;\;\;\;\;\;\omega_1=\mu B\cos\Theta
\label{2.631}
\ee

It is worth to note that the form (\ref{2.62}) of the considered transformation
provides us immediately with the asymptotic forms of the amplitudes
$a_1(s,T)$ and $a_2(s,T)$ for $s\to-\infty$ since the amplitudes $a_{\pm}(s,T)$ can take
arbitrary values $a_{\pm}(-\infty,T)$ in this limit satisfying only the condition
$|a_+(-\infty,T)|^2+|a_-(-\infty,T)|^2=1$. Namely we have simply in this limit
\be
a_1(s,T)\sim e^{-i\int_0^s(\dot\phi+\mu TB)\sin^2\frac{\Theta}{2}d\sigma}
\cos\frac{\Theta}{2}a_+(-\infty,T)+\nn
\\e^{-i\int_0^s(\dot\phi-\mu TB)\cos^2\frac{\Theta}{2}d\sigma}
\sin\frac{\Theta}{2}a_-(-\infty,T)\nn
\label{2.621}
\\
\\a_2(s,T)\sim e^{i\int_0^s(\dot\phi-\mu TB)\cos^2\frac{\Theta}{2}d\sigma+
i\phi(0)}\sin\frac{\Theta}{2}a_+(-\infty,T)-\nn
\\e^{i\int_0^s(\dot\phi+\mu TB)\sin^2\frac{\Theta}{2}d\sigma+
i\phi(0)}\cos\frac{\Theta}{2}a_-(-\infty,T)\nn
\ee

However since we are going to consider the case $a_+(-\infty,T)=1$ and
$a_-(-\infty,T)=0$ then for this case we get for $s\to-\infty$
\be
a_1(s,T)\sim e^{-i\int_0^s(\dot\phi+\mu TB)\sin^2\frac{\Theta}{2}d\sigma}
\cos\frac{\Theta}{2}\nn
\\\nn
\\a_2(s,T)\sim e^{i\int_0^s(\dot\phi-\mu TB)\cos^2\frac{\Theta}{2}d\sigma+i\phi(0)}
\sin\frac{\Theta}{2}
\label{2.64}
\ee

We can express further the amplitude $a_-(s,T)$, in which we are interested, by the $a_{1,2}$
ones inverting the transformation (\ref{2.62}) to get
\be
a_-(s,T)=e^{i\int_0^s(\dot\phi-\mu TB)\cos^2\frac{\Theta}{2}d\sigma}
\sin\frac{\Theta}{2}a_1(s,T)+\nn
\\e^{-i\int_0^s(\dot\phi+\mu TB)\sin^2\frac{\Theta}{2}d\sigma-
i\phi(0)}\cos\frac{\Theta}{2}a_2(s,T)
\label{2.65}
\ee

Moreover we can always assume that a limiting value of the field ${\bf B}(sT,T)$ for
$s\to+\infty$ coincides with its $z$-component to be
${\bf B}_+={\bf B}(+\infty,T)=(0,0,B_+)$, so that $\Theta(+\infty,T)=0$. Therefore
in the limit considered we get from (\ref{2.65})
\be
a_-(+\infty,T)=
\lim_{s\to+\infty}e^{-i\int_0^s(\dot\phi+\mu TB)\sin^2\frac{\Theta}{2}d\sigma-
i\phi(0)}a_2(s,T)
\label{2.66}
\ee

Consequently it is the amplitude $a_2(s,T)$ for which the above limit we have to consider.

The system (\ref{2.63}) can be rewritten further as the
     following linear system of second order equations
\begin{eqnarray}
\ddot a_1 -\left(\frac{\dot c_1^*}{c_1^*}+iT\omega_1\right)\dot a_1 +|c_1|^2a_1=0
\nn
\\\nn
\\\ddot a_2 -\left(\frac{\dot c_1}{c_1}-iT\omega_1\right)\dot a_2
  +|c_1|^2a_2=0
\label{2.7}
\end{eqnarray}
where the amplitudes $a_{1,2}$ decouple from each other
being however still related by (\ref{2.5}).

     By the following transformations
\begin{eqnarray}
a_1(s,T)=e^{\frac{1}{2}\int_0^s\left(\frac{\dot c_1^*}{c_1^*}+iT\omega_1
  \right)d\xi}b_1(s,T)\nn\\
a_2(s,T)=e^{\frac{1}{2}\int_0^s\left(\frac{\dot c_1}{c_1}-iT\omega_1
  \right)d\xi}b_2(s,T)
\label{2.8}
\end{eqnarray}
we bring the equations (\ref{2.7}) to Schr\"odinger types
\begin{eqnarray}
\ddot b_{1,2}(s,T)+T^2q_{1,2}(s,T)b_{1,2}(s,T)=0
\label{2.9}
\end{eqnarray}
where
\begin{eqnarray}
q_2(s,T)=-\frac{1}{4T^2}\left(\frac{\dot
      c_1}{c_1}-iT\omega_1\right)^2+|c_1|^2+\frac{1}{2T^2}\left(\frac{\dot
      c_1}{c_1}-i T\omega_1\right)^\cdot
\label{2.10}
\end{eqnarray}
while (for real $s$ and $T$) we have
\begin{eqnarray}
q_1(s,T)=q_2^*(s,T)
\label{2.11}
\end{eqnarray}

A dependence of the function
     $q_2(s,T)$ on $T$ is given by
\begin{eqnarray}
q_2(s,T)=\frac{1}{4}\mu^2B^2-\frac{i\mu B_z}{2T}\left(\frac{\dot B_z}{B_z}-\frac{\dot
  c_1}{c_1}\right)+\frac{1}{2T^2}\left[\left(\frac{\dot
  c_1}{c_1}\right)^\cdot-\frac{1}{2}\left(\frac{\dot
  c_1}{c_1}\right)^2\right]
\label{2.12}
\end{eqnarray}
where a dependence of $B, B_z, c_1$ on $T$ in (\ref{2.12}) is also
anticipated.
By (\ref{2.11}) we get a
corresponding dependence of $q_1(s,T)$ on $T$.

The equations (\ref{2.9}) are now basic for our further analysis since their form is just of the
stationary 1-D Schr\"odinger equation.

Taking into account (\ref{2.1}) and (\ref{2.6}) it is easy to check
that the last formula provides us with the following type of asymptotic behavior of
$q_{1.2}(s,T)$ for large $T$:
\begin{eqnarray}
q_{1,2}(s,T)=q_{1,2}^{(0)}(s)+\frac{1}{T}q_{1,2}^{(1)}(s)+\frac{1}{T^2}q_{1,2}^{(2)}(s)+\dots
\label{2.13}
\end{eqnarray}

Therefore the above form of dependence of $q_{1,2}(s,T)$ on $T$
permits us to apply to the considered case the method of fundamental
solutions. For this reason we shall start the next section with a
review of basic principles of the method suitably adapted to the
considered case.

\section{Fundamental solutions and their properties}

\hskip+2em Consider first $q_{1,2}(s,T)$ as functions of $s$. They are defined completely by an $s$-dependence
of field ${\bf B}(Ts,T)$. According to our assumptions, the latter is meromorphic on some Riemann
surface ${\bf R}_B(T)$. However, by (\ref{2.12}), $q_{1,2}(s,T)$ are
algebraic functions of ${\bf B}$,  ${\bf {\dot B}}$ and ${\bf {\ddot B}}$  and,
therefore, they are also meromorphic functions of $s$ defined again on some other Riemann
surfaces ${\bf R}_{1,2}$ determined by these algebraic dependencies. As it follows from (\ref{2.12}) topological structures
of ${\bf R}_{1,2}$ can be quite complicated. However, in what follows, we are interested in the
adiabatic limit $T \to +\infty$ by which structures of
${\bf R}_{1,2}$ should be determined for $T$ large
enough basically by the first term $q_{1,2}^{(0)}(s)$ of the expansion (\ref{2.13}).
In consequence, by (\ref{2.12}), it should be
determined by $\mu{\bf B}^{(0)}(s)$ i.e. by the first term of the
expansion (\ref{2.1}). Structures of ${\bf R}_{1,2}$ can turn out
to be much simpler in this limit.

     Despite this supposed complexity of $q_{1,2}(s,T)$  and of their Riemann surfaces we shall
introduce and discuss the fundamental solutions to the equations (\ref{2.9}) without simplifications.
We shall do it for the $q_2(s,T)$ case). An extension of the
discussion to the $q_1(s,T)$ case will
be obvious.

     A standard way of introducing the fundamental solutions is a construction of a Stokes graph
\cite{8,9,10} related to a given $q_2(s,T)$. Such a construction, according to Fr\"oman and Fr\"oman \cite{8}
and Fedoriuk \cite{9}, can be performed in the following way \cite{10}.

     Let $Z$ denote a set of all the points of ${\bf R}_2$ at which $q_2(s,T)$ has
     its single or double poles.
Let $\delta(x)$ be a meromorphic function on ${\bf R}_2$, the unique singularities of which are double poles at
the points collected by $Z$ with coefficients at all the poles equal to $1/4$ each.
(In a case when ${\bf R}_2$ is simply a complex plain the latter
function can be constructed in general with the
help of the Mittag-Leffler theorem \cite{17}. But for a case of
branched ${\bf R}_2$ the general procedure
is unknown to us).
     Consider now a function
\begin{eqnarray}
\tilde{q}_2(s,T)=q_2(s,T)+\frac{1}{T^2}\delta(s)
\label{3.1}
\end{eqnarray}

     The presence and the role of the $\delta$-term in (\ref{3.1}) are explained
     below. This term contributes
to (\ref{3.1}) if and only when the corresponding 'potential' function $q_2(s,T)$
contains simple or
second order poles. (Otherwise the corresponding $\delta$-term is put to zero).
It is called the Langer
term \cite{10,18}.

     The Stokes graph corresponding to the function $\tilde{q}_2(s,T)$ consists now
     of Stokes lines
emerging from roots (turning points) of $\tilde{q}_2(s,T)$. Stokes lines satisfy one
of the following equations:
\begin{eqnarray}
\Im \int_{s_i}^s\sqrt{\tilde{q}_2(\xi,T)}d \xi=0
\label{3.2}
\end{eqnarray}
with $s_i$ being a root of $\tilde{q}_2(s,T)$. We shall assume further a generic
situation when all roots
$s_i$ are simple.

     Stokes lines which are not closed end at these points of ${\bf R}_2$ (i.e. have
     the latter points
as their boundaries) for which the action integral in (\ref{3.2}) becomes infinite.
Of course such points
are singular for $\tilde{q}_2(s,T)$ and they can be its finite poles or its poles
lying at an infinity.

     Each such a singularity $z_k$ of $\tilde{q}_2(s,T)$ defines a domain called a
     sector. This is the
connected domain of  ${\bf R}_2$ bounded by Stokes lines and $z_k$
itself. The latter is also
a boundary for the Stokes lines or being an isolated boundary point of the sector
(as it is in the
case of the second order pole).

     In each sector the LHS in (\ref{3.2}) is only positive or only negative.

     Consider now equation (\ref{2.9}) for $b_2(s,T)$. Following Fr\"oman and Fr\"oman in each sector $S_k$ having a singular point $z_k$ at its
boundary one can
define a solution of the form:
\begin{eqnarray}
b_{2,k}(s,T) = \tilde{q}_2^{-\frac{1}{4}}(s,T){\cdot}
e^{\sigma i T W(s,T)}\chi_{2,k}(s,T) &
& k = 1,2,\ldots
\label{3.3}
\end{eqnarray}
where
\begin{eqnarray}
\chi_{2,k}(s,T) = 1 + \sum_{n{\geq}1}
\left( -\frac{\sigma}{2iT} \right)^{n} \int_{z_k}^{s}d{\xi_{1}}
\int_{z_k}^{\xi_{1}}d{\xi_{2}} \ldots
\int_{z_k}^{\xi_{n-1}}d{\xi_{n}}
\Omega(\xi_{1})\Omega(\xi_{2}) \ldots \Omega(\xi_{n})\times\nn\\
\\
\left( 1 - e^{-2\sigma iT{(W(s)-W(\xi_{1}))}} \right)
\left(1 - e^{-2\sigma iT{(W(\xi_{1})-W(\xi_{2}))}} \right)
\cdots
\left(1 - e^{-2\sigma iT{(W(\xi_{n-1})-W(\xi_{n}))}}
\right)\nn
\label{3.4}
\end{eqnarray}
with
\begin{eqnarray}
\Omega(s,T) = \frac{\delta(s)}{\tilde{q}_2^{\frac{1}{2}}(s,T)} -
{\frac{1}{4}}{\frac{\tilde{q}_2^{\prime\prime}(s,T)}
{\tilde{q}_2^{\frac{3}{2}}(s,T)}} +
{\frac{5}{16}}{\frac{\tilde{q}_2^{\prime 2}(s,T)}
{\tilde{q}_2^{\frac{5}{2}}(s,T)}}
\label{3.5}
\end{eqnarray}
and
\begin{eqnarray}
W(s,T) = \int_{s_{i}}^{s} \sqrt{\tilde{q}(\xi,T)}d\xi
\label{3.6}
\end{eqnarray}
where $s_i$ is a root of $\tilde{q}(s,T)$  lying at the boundary of $S_k$.

     In (\ref{3.3}) and (\ref{3.4}) a sign of $\sigma$ (=$\pm 1$) and an integration
     path are chosen in such a way
to have:
\begin{eqnarray}
\sigma \Im \left(W(\xi_{j}) - W(\xi_{j+1}) \right) \leq 0
\label{3.7}
\end{eqnarray}
for any ordered pair of integration variables (with $\xi_0=s$). Such
an integration path is then
called canonical. Of course, the condition (\ref{3.7}) means that $b_{2,k}(s,T)$
vanishes in its sector when
$s\to z_k$ along the canonical path.
     The Langer $\delta$-term appearing in (\ref{3.1}) and (\ref{3.5}) is necessary
     to ensure all the integrals
in (\ref{3.4}) to converge when $z_k$ is a first or a second order pole of
$\tilde{q}_2(s,T)$ or when the solutions
(\ref{3.3}) are to be continued to such poles. As it follows from (\ref{3.5}) each
such pole $z_k$ demands
a contribution to $\delta(s)$ of the form $\left(2(s-z_k)\right)^{-2}$, what has been
 already assumed in the corresponding
construction of $\delta(s)$.

It is now necessary to mention the main property of the fundamental solution method
which is that analytic continuations of fundamental solutions along canonical paths
ensures an immediate
pass to adiabatic limit on every stage of calculations performed with their use. This
property can be always utilized if all zeros of $\tilde q$-functions are simple and
distributions of their zeros and poles are discrete i.e. there are no accomodation
points for these singularities. We shall assume in the remainder that the two level
energy systems we are going to consider will satisfy the last conditions.

\section{The adiabatic limit in the fundamental solution approach}
\hskip+2em     Consider now the consequences of taking the large-$T$ limit for the
above description. We
assume that for a given $\tilde{q}_2(s,T)$ and its Riemann surface  ${\bf R}_2$ the
corresponding Stokes graph ${\bf G}_2$
is drawn. It is drawn, of course, on the Riemann surface $\sqrt{{\bf R}_2}$
corresponding to $\sqrt{\tilde{q}_2(s,T)}$.

     First let us notice that singular points of $\tilde{q}_2(s,T)$ such as its
     branch points and poles
depend in general on $T$. For both kinds of these singularities this
also means a dependence
on $T$ of jumps of $\tilde{q}_2(s,T)$ on its cuts as well as a $T$-dependence of
coefficients of its poles.

     According to the property ${\bf 5}^0$ of the magnetic field ${\bf B}$
     (see Sec. 2) we can expect that a
singular structure of $\tilde{q}_2(s,T)$, i.e. positions of its roots and poles, as
well as cut jumps and
pole coefficients, change smoothly in this limit to their final positions and values
respectively.
This limit structure is defined by a singularity structure of
$\tilde{q}_2^{(0)}(s,T)$ (see expansion
(\ref{2.13})). Therefore, both a topology of $\sqrt{{\bf R}_2}$ and the associated
Stokes graph ${\bf G}_2$ changes
accordingly to coincide eventually with the Riemann surface
$\sqrt{{\bf R}_2^{(0)}}$ and with the Stokes graph
${\bf G}_2^{(0)}$ corresponding to $\sqrt{\tilde{q}_2^{(0)}(s,T)}$ . This limiting
structure can be achieved in the following ways:

     {\bf a}. some of branch points and poles of
     $\tilde{q}_2(s,T)$ escape to infinities of ${\bf R}_2$;

      {\bf b}. some of branch points and poles of $\tilde{q}_+(s,T)$ approach the respective singularities
     of $\tilde{q}_2^{(0)}(s,T)$;

     {\bf c}. some of branch points and poles of $\tilde{q}_2(s,T)$ disappear because
      their respective
     jumps and coefficients vanish in the limit $T\to  +\infty$.

     Being more specific we expect that for $T$ large enough a set ${\bf S}_2$ of all
     singular
points of $\tilde{q}_2(s,T)$ (i.e. containing all its branch points and poles) consists
of three well
separated subsets ${\bf S}_2^{inf}$, ${\bf S}_2^{van}$ and ${\bf
  S}_2^{fin}$. The points of ${\bf S}_2^{inf}$ run to infinities of
${\bf R}_2$ when $T\to +\infty$.
Those of ${\bf S}_2^{van}$ disappear in this limit while those of ${\bf
  S}_2^{fin}$ coincide in this limit with the set ${\bf S}_2^{(0)}$
of singular points of $\tilde{q}_2^{(0)}(s,T)$ .

     Let us remove the
     points contained in ${\bf S}_2^{inf}\cup{\bf S}_2^{van}$ from the Riemann
     surface ${\bf R}_2$ ,
i.e. let us consider these points as regular for
     $\tilde{q}_2(s,T)$. Then ${\bf R}_2$ will transform
     into ${\bf R}_2^{fin}$ - a
Riemann surface which singular points coincide with those of the set ${\bf
  S}_2^{fin}$.

     Together with the previous operation let us remove from $\sqrt{{\bf R}_2}$ also
     the Stokes lines
generated by the points of ${\bf S}_2^{inf}\cup{\bf S}_2^{van}$ so that the remaining
 Stokes lines can be uniquely continued
to form the Stokes graph ${\bf G}_2^{fin}$ generated by the set ${\bf S}_2^{fin}$.
It is clear that the graph ${\bf G}_2^{fin}$ coincides
with ${\bf G}_2^{(0)}$ in the limit $T\to +\infty$.

     The above two operations will be called the adiabatic limit reduction or simply
     the reduction operation.

     As we have mentioned earlier there is a set of sectors and a corresponding set
     of
fundamental solutions defined in them associated with the graph  ${\bf G}_2$. By the
reduction operation
both sets can be reduced i.e. under this operation some sectors of
${\bf G}_2$ transform into corresponding sectors of ${\bf G}_2^{fin}$ whereas the
others disappear. Obviously, the latter sectors are those
which disappear when the limit $T\to +\infty$ is taken.

A good illustration for the above discussion can be an example considered in Sec.7,
namely the Nikitin model of the atom-atom scattering, for which the corresponding
rescaled {\bf B}-field is the following ${\bf B}(sT,T)=
\left(\left(b^2+s^2\right)^{-\frac{3}{2}},0,1\right)\frac{\Delta\epsilon}{\mu}$.
We can write for this case the corresponding Schr\"odinger equation (\ref{2.9}) using
the amplitudes $a_{\pm}(s,T)$ for which the respective $q_{\pm}(s,T)$ are following
\begin{eqnarray}
q_{\pm}(s,T)=\left[\frac{\Delta\epsilon}{2}\left(1+\frac{1}{\left(b^2+s^2\right)^3}
\right)^{\frac{1}{2}}\pm\frac{i}{2T}\left(\frac{6s(b^2+s^2)^2}{1+(b^2+s^2)^3}-
\frac{s}{b^2+s^2}-\frac{1}{s}\right)\right]^2
-\nn\\
\label{3.61}\\
\frac{3}{2}\frac{i\Delta\epsilon}{T}\frac{s}{\left(1+(b^2+s^2)^3\right)^
\frac{1}{2}(b^2+s^2)^\frac{5}{2}}-\hspace{40mm}\nn\\
\nn\\
\frac{1}{2T^2}\left[\frac{2s^2+b^2(b^2+s^2)}{s^2(b^2+s^2)}-
\frac{3}{2}\;\frac{\;4(b^2+s^2)^4(s^2-b^2)-4(b^2+s^2)(b^2+5s^2)+3s^2(b^2+s^2)}
{\left(1+(b^2+s^2)^3\right)^2}\right]\nn
\end{eqnarray}

     Equations (\ref{3.61}) show that in the limit $T\to + \infty$ the Stokes graph
     for the considered
problem is determined by the function
\begin{eqnarray}
q^{(0)}(s,T)=\frac{(\Delta\epsilon)^2}{4}\left(1+\frac{1}{\left(b^2+s^2\right)^3}
\right)
\label{3.62}
\end{eqnarray}

The graph is shown on Fig.1.

While each of $q_{\pm}(s,T)$ has 40 roots, five branch points at
  $s=\pm ib$ and at $s=s_k=\pm \left(e^{\frac{(2k+1)\pi i}{3}}-b^2\right)^
  {\frac{1}{2}}$
, $k=1,2,3$, as well as two poles at $s=0$, there are only six roots at
$s=s_k$, $k=1,2,3$ and only two poles at $s=\pm ib$ for $q^{(0)}(s,T)$.

     The functions $q_{\pm}(s,T)$ are determined on two sheeted
     Riemann surfaces ${\bf R}_{\pm}$ respectively
with the branch points at $s=\pm ib$ and at $s=s_k$, $k=1,2,3$ and with $40$ roots
distributed into halves on each
sheet of the surfaces. Therefore the Riemann surfaces $\sqrt
     {\bf R}_{\pm}$ corresponding to $\sqrt{q_\pm(s,T)}$
are four-sheeted with these $40$ roots being square root branch
points on them. When $T\to +\infty$ only six of these branch points survive
coinciding with
the six roots of $q^{(0)}(s,T)$ at $s=\pm s_k$, k=1,2,3 whereas ${\bf R}_{\pm}$
transforms into the complex $s$-plane since
the branch points of $q_{\pm}(s,T)$ at $s=\pm ib$ disappear, being transformed into
the second order poles
of $q^{(0)}(s,T)$. It is easy to check however that for finite but large $T$ these
six roots of $q^{(0)}(s,T)$ are
each split initially into two as. The split is the result of the square root branch
points at $s=\pm ib$ to which the
recovering of the finite $T$ transforms the poles of $q^{(0)}(s,T)$ at the same
points. The two copies
of each of these six roots lie of course on different sheets of ${\bf R}_{\pm}$.
Next, each of these 12 roots
is still split into three by the same reason of finiteness of $T$. In this way, on
each of
the two sheets of ${\bf R}_{\pm}$ there are $36$ roots grouped by
     three around their limit $s=\pm s_k$,
$k=1,2,3$ achieved for $T\to +\infty$.

     The remaining four roots of $q_{\pm}(s,T)$ are displaced in two pairs, one pair
      on each sheet
of ${\bf R}_{\pm}$, close to the points $s=0$ at which the second order poles of
$q_{\pm}(s,T)$ are localized. When
$T\to +\infty$ the roots in each pair collapse into $s=0$ multiplying the
corresponding second order
poles and thus causing mutual cancellations of the latter and
themselves in this limit.

Now let us focus our attention on
     the Stokes graph $\bf G _-$
generated by $q_-(s,T)$ on the first sheet of ${\bf R}_-$ as well as on the remaining
ones. It looks as in Fig.2.
(The Stokes graph $\bf G _+$ corresponding to $q_+(s,T)$ can be obtained from
$\bf G _-$ by complex
conjugation of the latter.) On the figure the wavy lines denote the cuts
corresponding to the branch points of the fundamental solutions
defined on ${\bf R}_-$. The sheet on Fig.2 cut along the wavy lines defines a domain
where all the
fundamental solutions $b_{-,1}(s,T),...,b_{-,\bar 2}(s,T)$ defined in
the corresponding sectors $S_1,...,S_{\bar 2}$ (shown
in the figure) are holomorphic.

     According to our earlier description of the behavior of the Riemann surface
     $\sqrt{R_+}$  when
$T\to +\infty$ the set ${\bf S} _-^{inf}$ corresponding to the considered case is
empty, ${\bf S} _-^{van}$ contains four points at
$s=0$ on each of the four sheets of $\sqrt{{\bf R}_-}$ (these four points correspond
to the second order poles
of $q_-(s,T)$) and the four branch points close to $s=0$, while ${\bf S}_-^{fin}$
contains all the remaining singular points of $\sqrt{q_-(s,T)}$.

\section{Systems which are not essentially different from their adiabatic limits
(NED systems)}

\hskip+2em The last example considered above shows us also that by changing
the amplitude representation to
the $a_{1,2}$ ones we get much simpler $s$-dependence for the corresponding functions
$q_{1,2}$ defining Eqs.(\ref{2.9}) and for their adiabatic limit $q_{1,2}^{(0)}$.
Namely for the interesting us amplitude $a_2$ we have
\be
q_2(s,T)=\frac{1}{4}(\Delta\epsilon)^2\left(1+\frac{1}{(b^2+s^2)^3}\right)-
         \frac{3\imath\Delta\epsilon}{2T}\frac{s}{b^2+s^2}-
         \frac{3}{4T^2}\frac{2b^2+s^2}{(b^2+s^2)^2}\nn\label{c}\\
\\q_2^{(0)}(s)=\frac{1}{4}(\Delta\epsilon)^2\left(1+\frac{1}{(b^2+s^2)^3}\right)\nn
\ee

It is seen from (\ref{c}) that both the functions $q_2(s,T)$ and $q_2^{(0)}(s)$ have
the same Riemann surfaces, namely the simple complex plain on which they have poles in
exactly the same points. They differ only by positions of their zeros the latter
being in a mutual one-to-one correspondence so that each zero of $q_2^{(0)}(s)$ is an
adiabatic limit of the corresponding zero of $q_2(s,T)$. Therefore the
Stokes graphs corresponding to both these functions are topologically equivalent having
the forms of Fig.1

As a consequence of this an application of the fundamental solution
method to the cases of Eqs.(\ref{2.9}) with respective $q_2(s,T)$ and $q_2^{(0)}(s)$
functions gives the same adiabatic limit for both these cases. We shall describe such
a situation as corresponding to a system which do not differ essentially from its
adiabatic limit and we shall call such a system the not-essentially-different one
(the NED-system).

It follows from the above discussion that the NED property is not an immanent one for
a system but can be achieved by choosing a suitable amplitude representation for a
system.
\vskip 12pt
\begin{tabular}{cc}
\psfig{figure=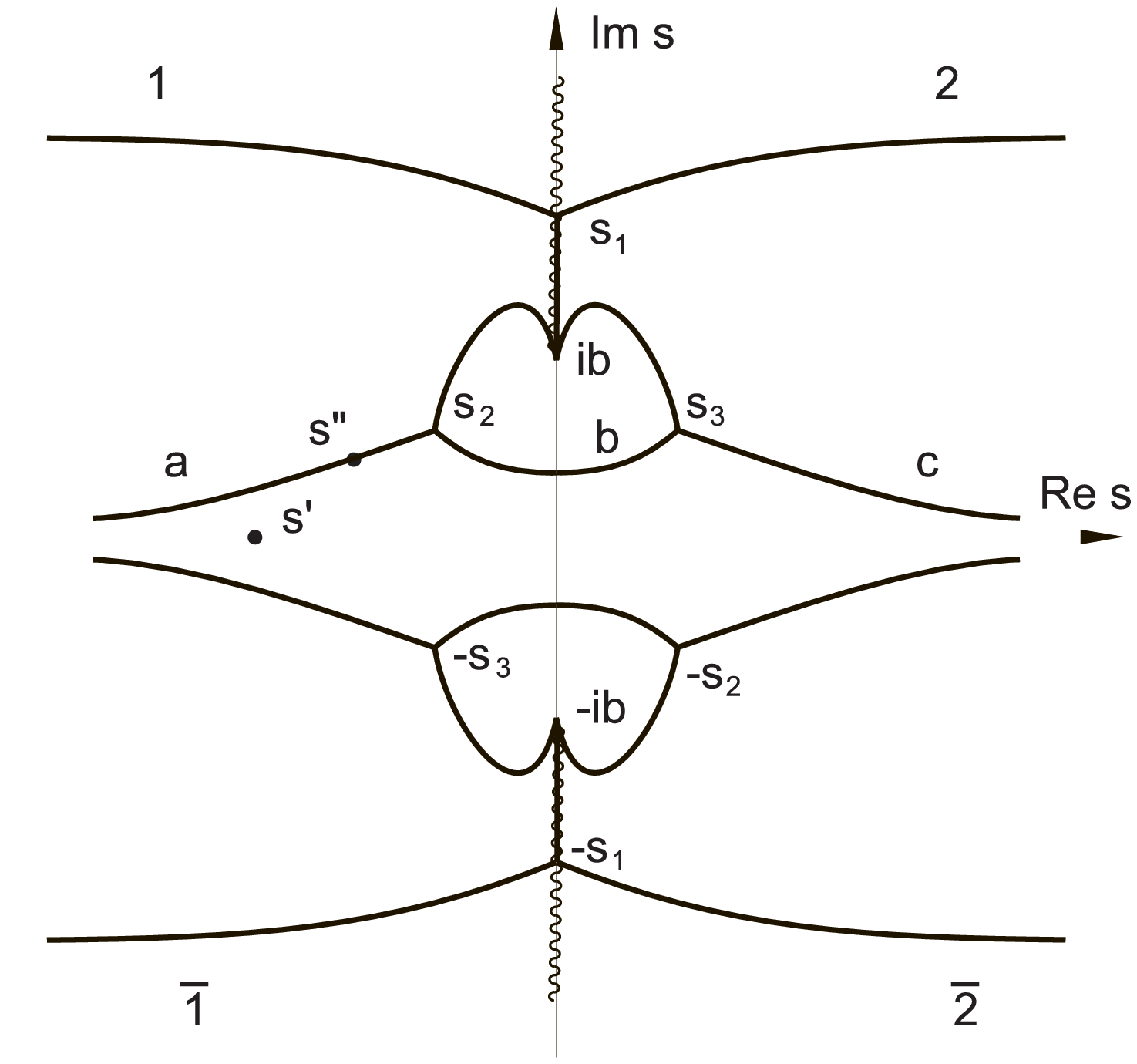,width=7cm}&\psfig{figure=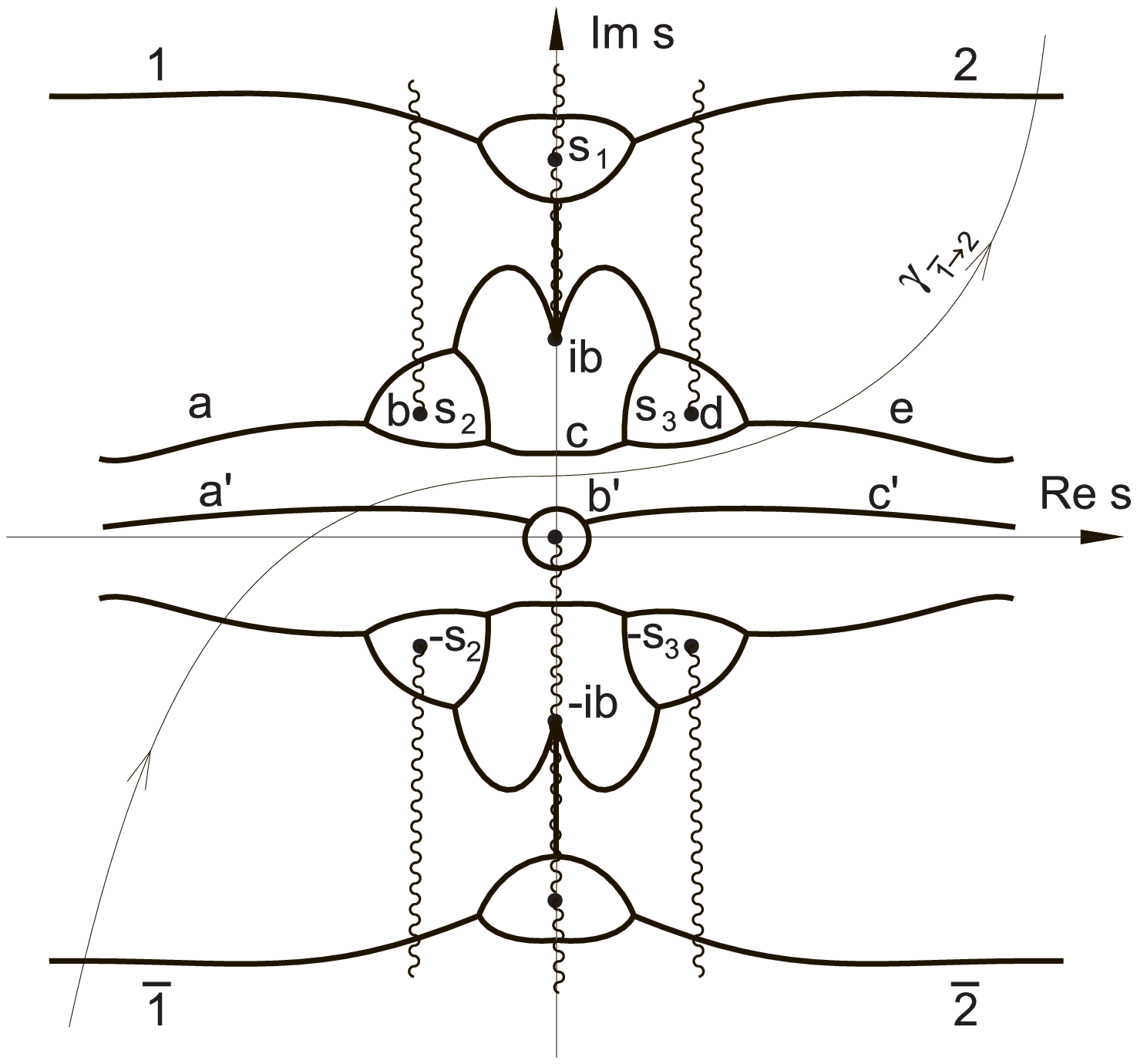,width=7cm} \\
Fig.1 The Stokes graph corresponding& Fig.2
The Stokes graph corresponding\\
      to $q_-^{(0)}(s)$ &       to $q_-(s,T)$
\end{tabular}
\section{Transition amplitudes for NED systems}

\hskip+2em As it follows from the discussion of the previous section systems with the
NED properties allow us for
as easy canonical continuations of fundamental solutions of interests as
they are for their corresponding adiabatically reduced forms. Therefore for such
systems we can consider them applying an exact procedure or using the adiabatical
limit for the latter to get correct results for adiabatical limit transition amplitudes.

We shall apply the procedure of canonical continuation of fundamental solutions
to the amplitude $a_2(s,T)$. First we have to express this amplitude by the fundamental
solutions and to satisfy the second of the conditions (\ref{2.64}). Canonically
continued to $-\infty$ with simple results of such continuations are the solutions
$b_{2,1}(s,T)$and $b_{2,\bar 1}(s,T)$ corresponding to the sectors $S_1$ and $S_{\bar 1}$
respectively shown in Fig.3 representing a Stokes graph corresponding to a
general NED system. We have
\begin{eqnarray}
a_2(s,T)=e^{\int_0^s\frac{1}{2}\left(\frac{\dot
      c_1}{c_1}-i\omega_1\right)(\sigma,T)d\sigma}(Ab_{2,\bar 1}(s,T)+Db_{2,1}(s,T))
\label{6.1}
\end{eqnarray}
where
\be
b_{2,\bar 1}(s,T)=q_2^{-\frac{1}{4}}(s,T)e^{-iT\int_{s_{\bar 1}}^s\sqrt{q_2(\xi,T)}d\xi}
\chi_{2,\bar 1}(s,T)\nn
\\b_{2,1}(s,T)=q_2^{-\frac{1}{4}}(s,T)e^{iT\int_{s_1}^s\sqrt{q_2(\xi,T)}d\xi}
\chi_{2,1}(s,T)
\label{6.11}
\ee
and where we have assumed the positive real value of $\sqrt{q_2(s,T)}$ on the physical
sheet. The lower integration limits in (\ref{6.11}) are respective zeros of
$q_2(s,T)$ shown in Fig.3.
\vskip 12pt
\begin{tabular}{c}
\psfig{figure=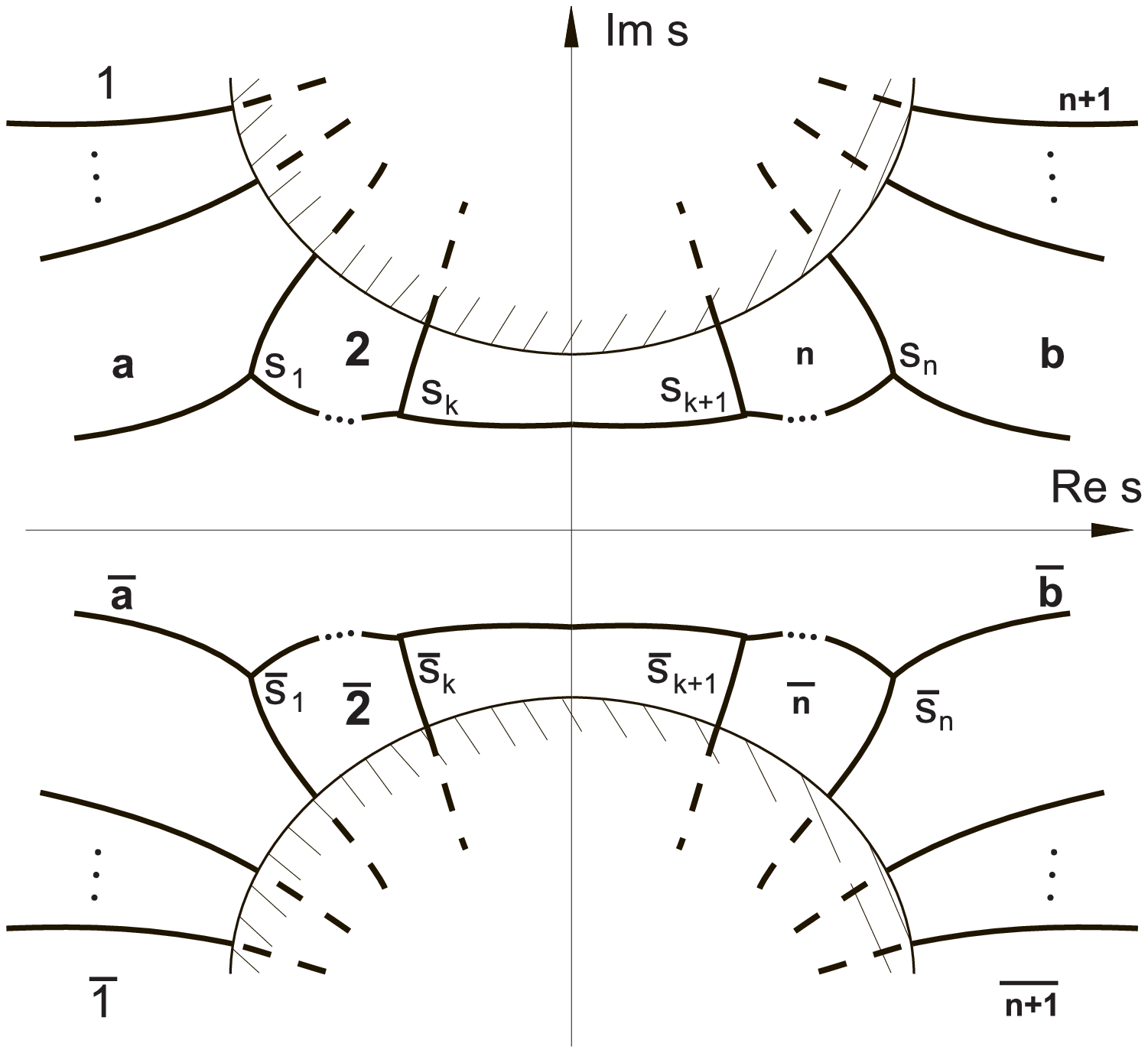,width=12cm}\\
Fig.3 The Stokes graph corresponding
to $q_2(s,T)$ of a NED system
\end{tabular}
\vskip 12pt
Taking into account that
\be
\frac{\dot c_1}{c_1}=\frac{\dot B}{B}+{\dot\Theta}\cot\Theta+i\dot\phi\nn
\\\nn
\\\frac{\dot B_z}{B}=\frac{\dot B}{B}\cos\Theta - {\dot\Theta}\sin\Theta
\label{6.111}
\ee
we get in the limits $s\to\pm\infty$ along the real axis
\be
iT\sqrt{q_2(s,T)}\sim\fr i\mu TB+\fr\frac{B_z}{B}\left(\frac{\dot B_z}{B_z}-
\frac{\dot c_1}{c_1}\right)=\fr i\mu TB-\frac{\dot\Theta}{2}\frac{1}{\sin\Theta}
-\fr i\dot\phi\cos\Theta
\label{6.112}
\ee
so that
\be
\frac{1}{2}\left(\frac{\dot
      c_1}{c_1}-i\omega_1\right)+iT\sqrt{q_2}\sim (i\dot\phi+i\mu TB)
      \sin^2\frac{\Theta}{2}+\fr\frac{\dot B}{B}-\frac{\dot\Theta}{2}
      \tan\frac{\Theta}{2}\nn
\\
\label{6.113}
\\\frac{1}{2}\left(\frac{\dot
      c_1}{c_1}-i\omega_1\right)-iT\sqrt{q_2}\sim (i\dot\phi-i\mu TB)
      \cos^2\frac{\Theta}{2}+\fr\frac{\dot B}{B}+\frac{\dot\Theta}{2}
      \cot\frac{\Theta}{2}\nn
\ee
in both the limits.

It is now easy to show that the Eqs.(\ref{6.1})-(\ref{6.113}) provide us with the
following asymptotic form of the amplitude $a_2(s,T)$ for $s\to-\infty$

\be
a_2(s,T)\sim \nn
\\\nn
\\A\left(\frac{\mu B_-}{2}\right)^{-\fr}\frac{1}
{\sin \frac{\Theta_0}{2}}e^{\int_0^{-\infty}\left[\frac{1}{2}\left(\frac{\dot
      c_1}{c_1}-i\omega_1\right)-iT\sqrt{q_2}-i(\dot\phi-\mu TB)
      \cos^2\frac{\Theta}{2}
      -\frac{\dot\Theta}{2}\cot\frac{\Theta}{2}\right]d\sigma
      -iT\int_{s_{\bar 1}}^0\sqrt{q_2}d\sigma
      -i\phi_0}\times\nn
\\\nn
\\e^{\int_0^si(\dot\phi-\mu TB)\cos^2\frac{\Theta}{2}d\sigma+i\phi_0}\sin\frac{\Theta}{2}+\nn
\ee
\be
\label{6.12}
\ee
\be
D\left(\frac{\mu B_-}{2}\right)^{-\fr}\frac{1}
{\cos \frac{\Theta_0}{2}}e^{\int_0^{-\infty}\left[\frac{1}{2}\left(\frac{\dot
      c_1}{c_1}-i\omega_1\right)+iT\sqrt{q_2}-i(\dot\phi+\mu TB)
      \sin^2\frac{\Theta}{2}
      -\frac{\dot\Theta}{2}\tan\frac{\Theta}{2}\right]d\sigma
      +iT\int_{s_1}^0\sqrt{q_2}d\sigma
      -i\phi_0}\times\nn
\\\nn
\\e^{\int_0^si(\dot\phi+\mu TB)\cos^2\frac{\Theta}{2}d\sigma+i\phi_0}\cos\frac{\Theta}{2}\nn
\ee
where $\Theta_0=\Theta(0)$, $\phi_0=\phi(0)$ and the infinite integrals in the
above formulae are finite.

Comparing now the formula (\ref{6.12}) with (\ref{2.621}) and (\ref{2.64}) we see that
we have to put $D=0$ in the formula (\ref{6.1}) while for the coefficient $A$
we get
\be
A=\left(\frac{\mu B_-}{2}\right)^{\fr}\sin \frac{\Theta_0}{2}
e^{\int_{-\infty}^0\left[\frac{1}{2}\left(\frac{\dot
      c_1}{c_1}-i\omega_1\right)-iT\sqrt{q_2}-i(\dot\phi-\mu TB)
      \cos^2\frac{\Theta}{2}
      -\frac{\dot\Theta}{2}\cot\frac{\Theta}{2}\right]d\sigma
      +iT\int_{s_{\bar 1}}^0\sqrt{q_2}d\sigma
      +i\phi_0}
\label{6.121}
\ee

Consequently it is the
solution $b_{2,\bar 1}(s,T)$ which will be continued canonically to the sectors
$n+1$ and
$\overline{n+1}$ from which it is subsequently continued to $+\infty$ of the real
$s$-axis. According to the figure this canonical continuation can be done by
representing $b_{2,\bar 1}(s,T)$ first as a linear combination of the next two
fundamental solutions $b_{2,2}(s,T)$ and $b_{2,\bar 2}(s,T)$ defined in the respective
sectors $2$ and $\bar 2$. Next the latter two solutions have to be expressed in
the same way by a pair of fundamental solutions of the sectors $3$ and $\bar 3$ and
so on up to the moment when the fundamental solutions of the sectors $n+1$ and
$\overline{n+1}$ enter this procedure. Representing the corresponding fundamental
solutions in the form
\be
b_{2,k}(s,T)=q_2^{-\frac{1}{4}}(s,T)e^{iT\int_{s_{k-1}}^s\sqrt{q_2(\xi,T)}d\xi}
\chi_{2,k}(s,T)\\
\label{2.14}
b_{2,\bar k}(s,T)=q_2^{-\frac{1}{4}}(s,T)e^{-iT\int_{s_{k-1}}^s\sqrt{q_2(\xi,T)}d\xi}
\chi_{2,\bar k}(s,T)\nn
\\k=2,3,\dots,n+1\nn
\ee
this chain of operations can be handled by the following multiplications of matrices
\be
M=M_1M_2\dots M_n\nn\\
\nn\\
M_1=\frac{1}{\chi_{2,2\to\bar 2}}
\left[\begin{array}{cc}
0&0\\
-i\alpha_{\bar 1,1}\chi_{2,\bar 1\to\bar 2}&\chi_{2,\bar 1\to 2}
\end{array}\right]\nn\\
\nn\\
\nn\\
M_k=\frac{1}{\chi_{2,k+1\to\overline{k+1}}}\left[\begin{array}{cc}
e^{\beta_k}\chi_{2,k\to\overline{k+1}} &i\alpha_{\bar k,k} e^{\beta_k}\chi_{2,k\to k+1}
\\-i\alpha_{\bar k,k} e^{\beta_{\bar k}}\chi_{2,\bar k\to\overline{k+1}}&e^{\beta_{\bar k}}
\chi_{2,\bar k\to k+1}
\end{array}\right]\\
\label{15}
\nn\\
\nn\\
\alpha_{\bar k,k}=e^{iT\int_{s_{\bar k}}^{s_k}\sqrt{q_2(s,T)}ds}\nn\\
\nn\\
\beta_{k+1}=iT\int_{s_k}^{s_{k+1}}\sqrt{q_2(s,T)}ds,\;\;
\beta_{\overline {k+1}}=-iT\int_{s_{\bar k}}^{s_{\overline{k+1}}}
\sqrt{q_2(s,T)}ds\nn\\
k=1,\dots,n\nn
\ee
so that
\be
b_{2,\bar 1}(s,T)=M_{21}b_{2,n+1}(s,T)+M_{22}b_{2,\overline{n+1}}(s,T)
\label{16}
\ee

Let us note that the phase integrals defining the coefficients $\beta_k$ and
$\beta_{\bar k},\;k=2,\;...\;n$, are purely imaginary. Moreover
the coefficients $\alpha_{\bar k,k},\;k=1,...,n$, become pure real and equal to each
other while the
coefficients $\beta_k$ become equal to
$-\beta_{\bar k},\;k=2,...,\;n$ in the adiabatic limit $T\to +\infty$. To be more
precise in these latter statements let $s_{\bar k}',\;k=1,...,n$, denote points where
the antiStokes line emanating from $s_{\bar k},\;k=1,...,n$, crosses the Stokes line
passing by the points $s_1,s_2,...,s_n$. Then by the assumption ${\bf 8}^0c$ of Sec.2
we have
\be
\int_{s_{\bar k}}^{s_k}\sqrt{q_2(s,T)}ds=\int_{s_{\bar k}}^{s_{\bar k}'}
\sqrt{q_2(s,T)}ds+\int_{s_{\bar k}'}^{s_k}\sqrt{q_2(s,T)}ds=\nn\\
\int_{s_{\bar k}}^{s_k'}\sqrt{q_2(s,T)}ds +O(\frac{1}{T})\nn\\
\nn\\
\int_{s_{\bar k}}^{s_k'}\sqrt{q_2(s,T)}ds=\int_{s_{\bar 1}}^{s_1'}\sqrt{q_2(s,T)}ds
\label{16a}\\
\nn\\
\int_{s_{\bar k}}^{s_{\overline{k+1}}}\sqrt{q_2(s,T)}ds=
\int_{s_{\bar k}'}^{s_{\overline{k+1}}'}\sqrt{q_2(s,T)}ds=\nn\\
\left(\int_{s_{\bar k}'}^{s_k}+\int_{s_k}^{s_{k+1}}+
\int_{s_{k+1}}^{s_{\overline{k+1}}'}\right)\sqrt{q_2(s,T)}ds=\nn\\
\int_{s_k}^{s_{k+1}}\sqrt{q_2(s,T)}ds+O(\frac{1}{T})\nn\\
k=1,...,n\nn
\end{eqnarray}
i.e. each point $s_{\bar k}',\;k=2,...,\;n$ tends to its corresponding limit
$s_k,\;k=2,...,n$ when $T\to+\infty$ with the rates shown in (\ref{16a}).

Rewriting the Eqs.(\ref{6.1}) as
\be
a_2(s,T)=Ae^{\fr\int_0^s\left(\frac{\dot c_1}{c_1}-iT\omega_1\right)d\xi}b_{2,\bar 1}(s,T)
\label{17}
\ee
and taking into account (\ref{2.66}) we get
\be
a_-(+\infty,T)=M_{21}\left(\frac{B_-}{B_+}\right)^{\fr}\sin\frac{\Theta_0}{2}\times\nn
\\\nn
\\\exp\left\{\int_{-\infty}^0\left[\frac{1}{2}\left(\frac{\dot
      c_1}{c_1}-i\omega_1\right)-iT\sqrt{q_2}-i(\dot\phi-\mu TB)
      \cos^2\frac{\Theta}{2}
      -\frac{\dot\Theta}{2}\cot\frac{\Theta}{2}\right]ds+\right.\nn
\\\nn
\\\int_0^{+\infty}\left[\fr\left(\frac{\dot c_1}{c_1}-iT\omega_1\right)+
iT\sqrt{q_2}-i(\dot\phi+\mu TB)\sin^2\frac{\Theta}{2}\right]ds+\nn
\\\nn
\\\left.iT\int_{s_{\bar 1}}^0\sqrt{q_2}d\sigma+iT\int_{s_n}^0\sqrt{q_2}ds\right\}
\label{18}
\ee
since the second term in (\ref{16}) vanishes in the limit $s\to+\infty$ along the
real axis (because $\Theta(+\infty,T)=0$ by assumption).

The formula (\ref{18}) is just the one which corrects the erroneuos formula (29) of
the paper \cite{01} (as well as the other formulae corresponding to other cases considered in the cited
 paper).

It should be stressed that the formula (\ref{18}) is {\it exact}. In this form
it looks however very complicate because
of the complicated structure of the matrix element $M_{21}$. The latter is
polynomial with respect to the coefficients $\alpha_{\bar k,k},\;k=1,...,n$,
and rational with respect to the $\chi$-coefficients. Exposing its linear terms
in $\alpha$'s we get for it
\be
M_{21}=-i\prod_{k=2}^{n}\chi_{\bar k\to k}^{-1}\sum_{k=1}^ne^{iT\int_{s_{\bar k}}^{s_k}
\sqrt{q_2(s,T)}ds-iT\int_{s_{\bar 1}}^{s_{\bar k}}\sqrt{q_2(s,T)}ds+
iT\int_{s_k}^{s_n}\sqrt{q_2(s,T)}ds}\times\nn\\
\chi_{\bar k\to \overline{k+1}}\prod_{l=1}^{k-1}
\chi_{\bar l\to l+1}\prod_{l=k+1}^n\chi_{l\to\overline{l+1}}+ \cdots
\label{18a}
\ee

Nevertheless in the adiabatic limit $T\to +\infty$ the formulae (\ref{18}) and
(\ref{18a})
simplify greatly since then all
$\chi$'s coefficients of $M_{21}$ become equal to $1$ and in the multiplication of
the limiting matrices $M_k$'s, $k=1,\dots,n$, all terms containing powers of the
factors $\alpha_{\bar k,k},\;k=1,...,n$
higher than the first ones have to be neglected. Further everywhere where $B(s,T)\ne 0$ the
adiabatic limits $T\to+\infty$ of $\fr\left(\frac{\dot c}{c}-iT\omega\right)\pm
iT\sqrt{q_2(s,T)}$ are exactly the same in their forms as those for $s\to\pm\infty$
given
by (\ref{6.113}). The condition $B(s,T)\ne 0$ is satisfied obviously for the
integrals in the formula (\ref{18}) taken along the real axis. However the
phase integrals
defining the element $M_{12}$ are taken between zeros of $\sqrt{q_2(s,T)}$ which in
the adiabatic limit coincide with the ones of $B_0(s)$. Nevertheless this trouble can
be easily avoided by representing the corresponding integrations linking pairs of zeros
($s_{\bar k},s_k$),
($s_{\bar 1},s_{\bar k}$) and ($s_k,s_n$), $k=1,...,n$, by the ones along closed
contours $C_{s_{\bar k}s_{k}}$, $C_{s_{\bar 1}s_{\bar k}}$ and $C_{s_ks_n}$,  surrounding respective
pairs of zeros. The same idea
applies to the two integrations between the pair of points ($s_{\bar 1},0$) and
($s_n,0$)
except that the corresponding contours $C_{s_{\bar 1}0}$ and $C_{s_n0}$  are not
closed but starts and ends at $s=0$ points lying on two different sheets of ${\bf R}_B$.
Therefore making first use of the Eqs.(\ref{16a}) we can apply the asymptotics
 (\ref{6.113}) to all the phase integrals in the formulae (\ref{18}) and (\ref{18a})
 so that the former takes
 the following form when the integrations along the real axis is performed
 \be
a_-^{ad}(+\infty,T)=-i\tan\frac{\Theta_0}{2}
e^{\frac{1}{4}\left(\int_{C_{s_{\bar 1}s_1}}+\int_{C_{s_{\bar 1}0}}+\int_{C_{s_n}0}\right)
\left(i\mu TB-\frac{\dot\Theta}{\sin\Theta}
-i \dot\phi\cos\Theta\right)ds}\times\nn
\\\sum_{k=1}^n
e^{-\frac{1}{4}\left(\int_{C_{s_1}s_{k}}-
\int_{C_{s_ks_{n}}}\right)\left(i\mu TB-\frac{\dot\Theta}{\sin\Theta}
-i\dot\phi\cos\Theta\right)ds}
\label{22}
\ee
where it is assumed that all the quantities are now determined by the asymptotic
field ${\bf B}_0$.

In the present form of the formula (\ref{22}) only the integrations of the middle term
$-\frac{\dot\Theta}{\sin\Theta}$ in the exponents can be performed explicitly
(since $\int \frac{\dot\Theta}{\sin\Theta}ds=\ln \tan\frac{\Theta}{2}$). First let us note that
because $\frac{\dot\Theta}{\sin\Theta}=
\fr\left(\frac{{\dot B}_0-{\dot B}_{0,z}}{B_0-B_{0,z}}-
\frac{{\dot B}_0+{\dot B}_{0,z}}{B_0+B_{0,z}}\right)$, so that $\ln \tan\frac{\Theta}{2}=
\fr\ln\frac{B_0-B_{0,z}}{B_0+B_{0,z}}$,
we can always choose all the integration contours in (\ref{22})
in such a way to avoid possible
roots of $B_{0,z}\pm B_0=0$ so that the unique singularities which remain inside these
contours are branch points of $B_0(s)$. Therefore to
the corresponding integrals of $-\frac{\dot\Theta}{\sin\Theta}$
along the closed contours $C_{s_1s_k},\;
C_{s_ks_n},\;k=1,...,n$, and $C_{s_{\bar 1}s_1}$  can contribute
only roots
of the function $F(s)\equiv\frac{B_0(s)-B_{0,z}(s)}{B_0(s)+B_{0,z}(s)}$. Net results of these contributions depends however
on detailes of mapping of the $s$-Riemann surface on the $F$-one. If after such a
mapping a closed contour $C_\gamma$ rounds the zero point of the $F$-plane
$n_\gamma$ times (we take $n_\gamma$ to be positive for anticlock orientation of the
contour and negative for the opposite case) then a contribution of this zero point
to the corresponding
contour integral of $-\frac{\dot\Theta}{\sin\Theta}$ is $-i\pi n_\gamma$.
The remaining two open integrals along the contours $C_{s_{\bar 1}0}$ and
$C_{s_n0}$ can contribute only by their limits giving
\be
-\frac{1}{4}\left(\int_{C_{s_{\bar 1}0}}+\int_{C_{{s_n}0}}\right)\frac{\dot\Theta}
{\sin\Theta}=-\ln\tan\frac{\Theta_0}{2} +il\frac{\pi}{2}
\label{221}
\ee
with some integer $l$ since $B_{0,z}(s)$ is regular at the points $s_{\bar 1},s_n$ and values of $B_0(s)$
on both the sheets differ by sign so that $F_2(0)=F_1^{-1}(0)$ where $F_{1,2}(0)$
 are values (both real) of $F(s)$ at $s=0$ on the 'first' and 'second' sheets respectively.

Therefore we obtain the following final result
\be
a_-^{ad}(+\infty,T)=
-i^{l+1}e^{-\frac{1}{4}i\pi n_{s_{\bar 1}s_1}}e^{\frac{1}{4}\left(\int_{C_{s_{\bar 1}s_1}}+\int_{C_{s_{\bar 1}0}}+\int_{C_{s_n}0}\right)
\left(i\mu TB-i \dot\phi\cos\Theta\right)ds}\times\nn
\\\sum_{k=1}^ne^{\frac{1}{4}i\pi( n_{s_1s_k}-n_{s_ks_n})}
e^{-\frac{1}{4}\left(\int_{C_{s_1}s_k}-
\int_{C_{s_ks_n}}\right)\left(i\mu TB-i\dot\phi\cos\Theta\right)ds}
\label{222}
\ee

Since $\dot\phi\cos\Theta=\frac{B_{0,z}}{B_0}\frac{B_{0,x}{\dot B}_{0,y}-
B_{0,y}{\dot B}_{0,x}}{B_0^2-B_{0,z}^2}$ we can shrink all the integrations in
(\ref{222}) to paths linking respective points to get
\be
a_-^{ad}(+\infty,T)=
-i^{l+1}e^{-\frac{1}{4}i\pi n_{s_{\bar 1}s_1}}e^{\frac{1}{2}\left(\int_{s_{\bar 1}}^{s_1}+\int_{s_{\bar 1}}^0+\int_{s_n}^0\right)
\left(i\mu TB-i \dot\phi\cos\Theta\right)ds}\times\nn\\
\sum_{k=1}^n
e^{\frac{1}{4}i\pi( n_{s_1s_k}-n_{s_ks_n})}
e^{-\frac{1}{2}\left(\int_{s_1}^{s_k}-
\int_{s_k}^{s_n}\right)\left(i\mu TB-i\dot\phi\cos\Theta\right)ds}
\label{223}
\ee

We should remember that all the integrations in (\ref{223}) run along paths avoiding
roots of the equations $B_{0,z}\pm B_0=0$.

For the corresponding transition probability we obtain
\be
P_-^{ad}(T)=e^{-\Im\int_{s_{\bar 1}}^{s_1}\left(\mu TB- \dot\phi\cos\Theta\right)ds}
\left|\sum_{k=1}^n
e^{\frac{1}{4}i\pi( n_{s_1s_k}-n_{s_ks_n})}
e^{-\frac{1}{2}\left(\int_{s_1}^{s_k}-
\int_{s_k}^{s_n}\right)\left(i\mu TB-i\dot\phi\cos\Theta\right)ds}\right|^2
\label{224}
\ee

Formulae similar to (\ref{223}) and (\ref{224}) have been found by Joye, Mileti and Pfister \cite{4}.
In fact if we apply the assumptions made in the last paper by its authors these formulae
become identical, up to an overall
phase in (\ref{223}), with the corresponding ones found by the authors mentioned.

The last formulae take on particularly simple forms for the case of two turning points
lying on the upper Stokes line drawn on Fig.3. when the equations $B_{0,z}\pm B_0=0$
have no solutions inside the strip bounded by the two Stokes lines on Fig.3 and on
the lines themselves. We can then deform all integration paths in the formula (\ref{223})
to ones along the Stokes and antyStokes lines so that the corresponding integrals
will have explicitly pure real or pure imaginary values. We get for this case
\be
a_-^{ad}(+\infty,T)=
-i^{l+1}e^{-\frac{1}{4}i\pi n_{s_{\bar 1}s_1}}
e^{\frac{1}{2}\left(\int_{s_{\bar 1}}^{s_1}+\int_{s_{\bar 1}}^0+\int_{s_2}^0\right)
\left(i\mu TB-i \dot\phi\cos\Theta\right)ds}\times\nn\\
\left(e^{\frac{1}{4}i\pi n_{s_1s_2}}
e^{-\frac{1}{2}\int_{s_1}^{s_2}\left(i\mu TB-i\dot\phi\cos\Theta\right)ds}+
e^{-\frac{1}{4}i\pi n_{s_1s_2}}
e^{\frac{1}{2}\int_{s_1}^{s_2}\left(i\mu TB-i\dot\phi\cos\Theta\right)ds}\right)=\nn\\
-2i^{l+1}e^{-\frac{1}{4}i\pi n_{s_{\bar 1}s_1}+
\frac{1}{2}i\Re\left(+\int_{s_{\bar 1}}^0+\int_{s_2}^0\right)
\left(\mu TB- \dot\phi\cos\Theta\right)ds}
e^{-\fr\Im\int_{s_{\bar 1}}^{s_1}\left(\mu TB- \dot\phi\cos\Theta\right)ds}\times\nn\\
\cos\left(\fr\Re\int_{s_1}^{s_2}\left(\mu TB- \dot\phi\cos\Theta\right)ds-
\frac{1}{4}\pi n_{s_1s_2}\right)
\label{225}
\ee
so that for the corresponding transition probability we get
\be
P_-^{ad}(+\infty,T)=e^{-\Im\int_{s_{\bar 1}}^{s_1}
\left(\mu TB- \dot\phi\cos\Theta\right)ds}
\cos^2\left(\fr\Re\int_{s_1}^{s_2}\left(\mu TB- \dot\phi\cos\Theta\right)ds-
\frac{1}{4}\pi n_{s_1s_2}\right)
\label{226}
\ee

\section{Examples of NED systems}

\hskip+2em An example of a class of fields $\textbf{B}$ with the NED property has been
considered recently by Berman {\it et al} \cite{20}. The fields are defined by
putting
$B_z(sT,T)=B_\infty,\;B_x(sT,T)=f(s)\cos(\omega_0 sT),\;B_y(sT,T)=f(s)
\sin(\omega_0 sT)$ with $f(s)$ having the properties ${\bf 1}^0-{\bf 3}^0$ of the
field ${\bf B}$ and vanishing at both infinities of the real axis. This problem is
however unitary equivalent to the one with the field ${\bf B}=[f(s),0,B_\infty-\frac{\omega_0}
{\mu}]$ so that for this case we have $B=\sqrt{\left(\frac{\Omega}{\mu}
\right)^2+f^2(s)}$ where $\Omega=\mu B_\infty-\omega_0$ and $\dot\phi\equiv 0$.

Assuming for $f(s)$ properties desired by the assumption ${\bf 1}^0-{\bf 8}^0$ of
Sec.2 we get using the formula (\ref{223})
\be
a_-^{ad}(+\infty,T)=-i^{l+1}e^{-\frac{1}{4}i\pi n_{s_{\bar 1}s_1}}e^{\frac{1}{2}i\mu T\left(\int_{s_{\bar 1}}^{s_1}+\int_{s_{\bar 1}}^0+\int_{s_n}^0\right)
\sqrt{\left(\frac{\Omega}{\mu}
\right)^2+f^2(s)}ds}\times\nn\\
\sum_{k=1}^ne^{\frac{1}{4}i\pi( n_{s_1s_k}-n_{s_ks_n})}
e^{-\frac{1}{2}i\mu T\left(\int_{s_1}^{s_k}-
\int_{s_k}^{s_n}\right)\sqrt{\left(\frac{\Omega}{\mu}
\right)^2+f^2(s)}ds}
\label{23}
\ee
where $s_{\bar 1}$ and $s_k,\;k=1,...,n$, are roots of the equations $f(s)=\pm i\frac{\Omega}{\mu}$.

If there are only two turning points $s_1$ and $s_2$ then according to formula
(\ref{225}) we get
\be
a_-^{ad}(+\infty,T)=-2i^{l+1}e^{-\frac{1}{4}i\pi n_{s_{\bar 1}s_1}+
\frac{1}{2}i\mu T\Re\left(+\int_{s_{\bar 1}}^0+\int_{s_2}^0\right)
\sqrt{\left(\frac{\Omega}{\mu}\right)^2+f^2(s)}ds}\times\nn\\
e^{-\fr\mu T\Im\int_{s_{\bar 1}}^{s_1}
\sqrt{\left(\frac{\Omega}{\mu}\right)^2+f^2(s)}ds}
\cos\left(\fr\mu T\Re\int_{s_1}^{s_2}
\sqrt{\left(\frac{\Omega}{\mu}\right)^2+f^2(s)}ds-
\frac{1}{4}\pi n_{s_1s_2}\right)
\label{231}
\ee

It is now not difficult to establish that the contour $C_{s_1s_2}$ rounds
the zero point on the $F$-plane twice (see Fig.4 and Fig.5). Therefore we obtain
finnally for this case
\be
a_-^{ad}(+\infty,T)=-2i^{l+1}e^{-\frac{1}{4}i\pi n_{s_{\bar 1}s_1}+
\frac{1}{2}i\mu T\Re\left(+\int_{s_{\bar 1}}^0+\int_{s_2}^0\right)
\sqrt{\left(\frac{\Omega}{\mu}\right)^2+f^2(s)}ds}\times\nn\\
e^{-\fr\mu T\Im\int_{s_{\bar 1}}^{s_1}
\sqrt{\left(\frac{\Omega}{\mu}\right)^2+f^2(s)}ds}
\sin\left(\fr\mu T\Re\int_{s_1}^{s_2}
\sqrt{\left(\frac{\Omega}{\mu}\right)^2+f^2(s)}ds\right)
\label{232}
\ee
and for the corresponding transition amplitude
\be
P_-^{ad}=4e^{-\mu T\Im\int_{s_{\bar 1}}^{s_1}
\sqrt{\left(\frac{\Omega}{\mu}\right)^2+f^2(s)}ds}
\sin^2\left(\fr\mu T\Re\int_{s_1}^{s_2}
\sqrt{\left(\frac{\Omega}{\mu}\right)^2+f^2(s)}ds\right)
\label{233}
\ee

The last two formulae have been obtained earlier by Nikitin and Umanskii
\cite{11} as well as by Crothers \cite{16} and by Davies and Pechukas \cite{19} using
the steepest-descent methods.
\vskip 12pt
\begin{tabular}{cc}
\psfig{figure=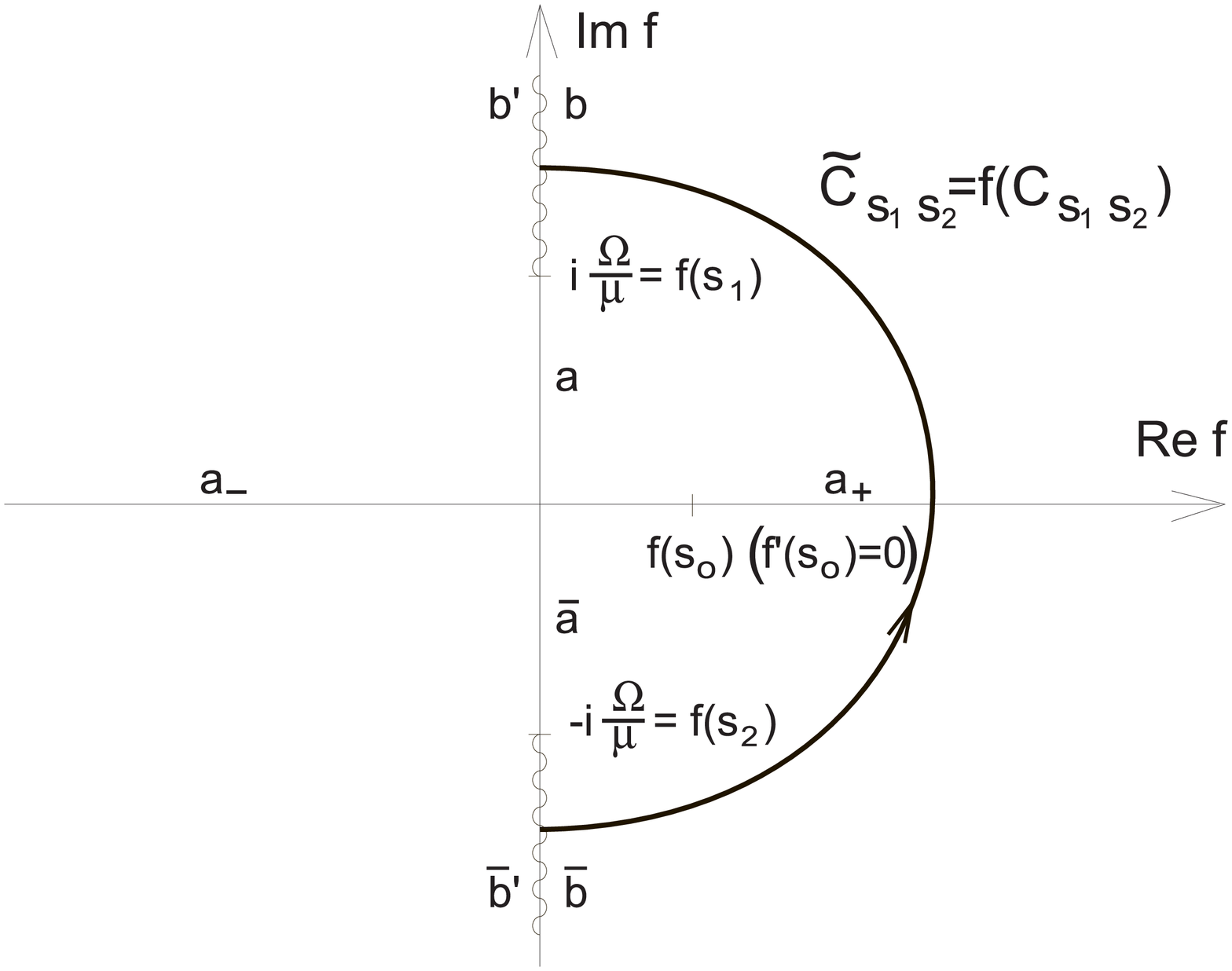,width=7cm}&\psfig{figure=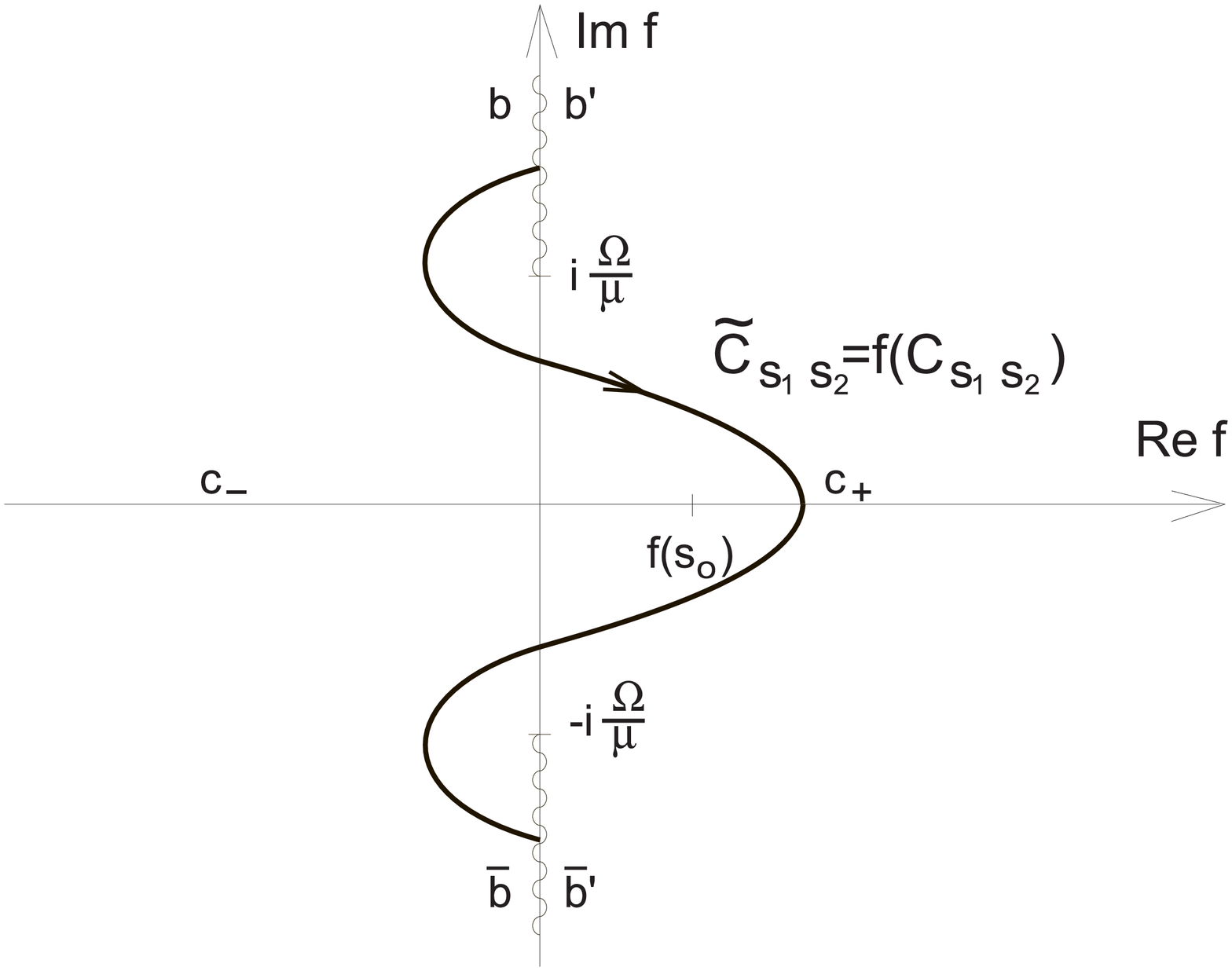,width=7cm}
\end{tabular}

Fig.4 The integration contour $C_{s_1s_2}$ mapped into the $f$-plane
\vskip 12pt
\begin{tabular}{cc}
\psfig{figure=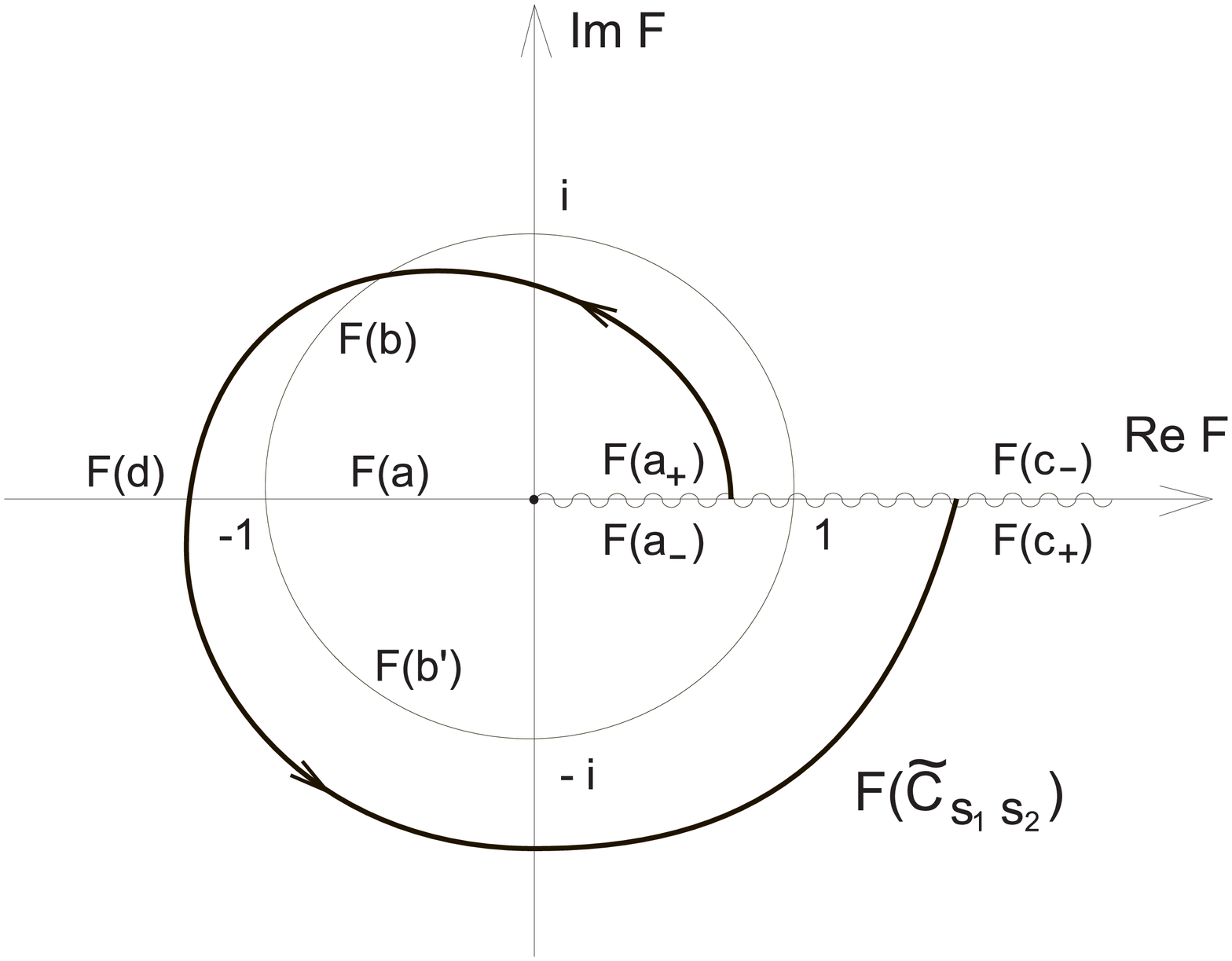,width=7cm}&\psfig{figure=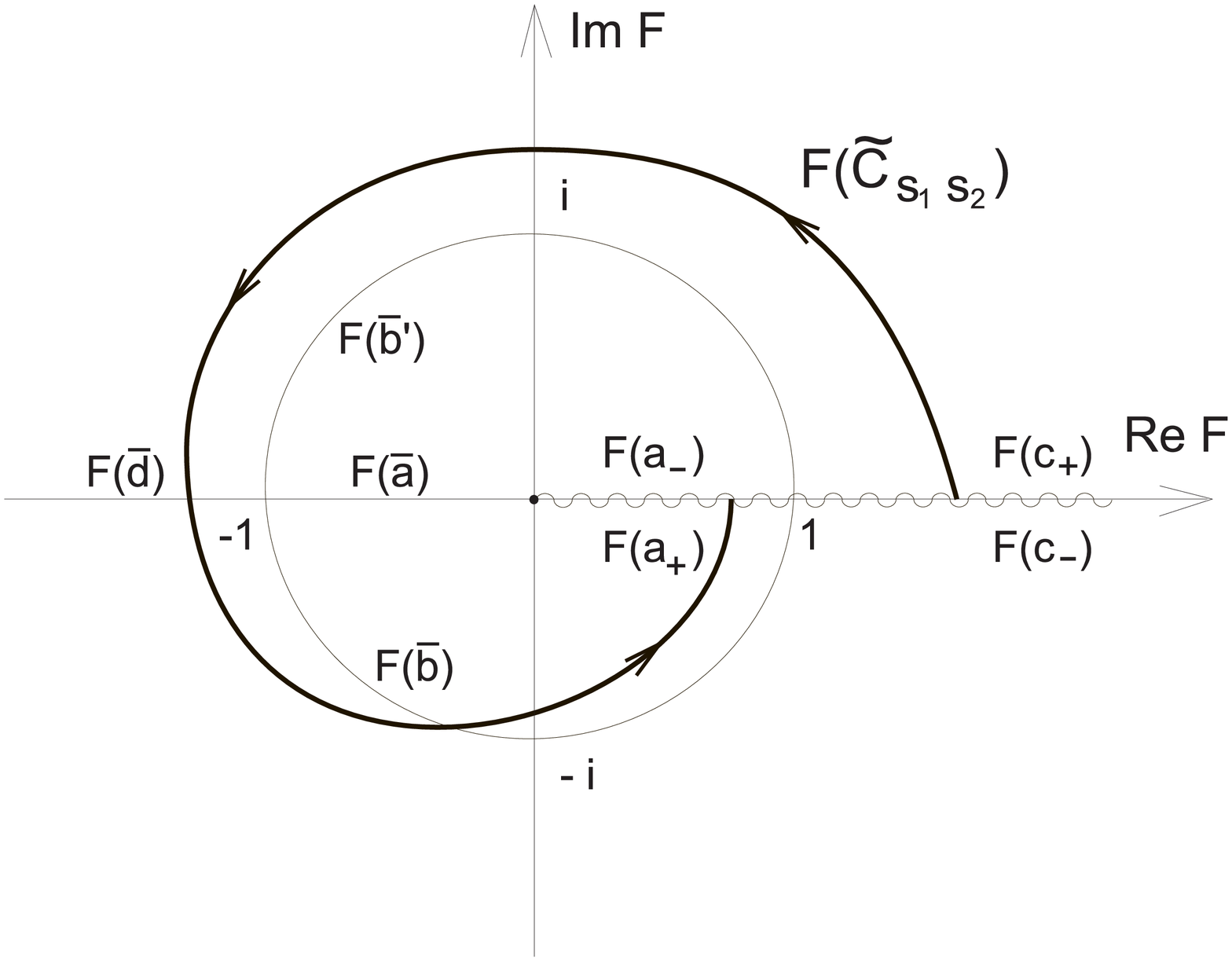,width=7cm}
\end{tabular}

Fig.5 The integration contour $C_{s_1s_2}$ mapped into the $F$-plane
\vskip 12pt

As a second example we shall consider again the Nikitin Hamiltonian for the atom-
atom scattering. The model of Nikitin \cite{12} describes the scattering
A*+B$\rightarrow$A+B+$\Delta\epsilon$ of the exited atom
A* moving with a small velocity $v$ with the impact parameter $b'$ and scattered by the atom
B. The interaction between the atoms is of the dipol-dipol type. The latter example was
analyzed in the context of the adiabatic limit $v\to0$ also by Joye {\it et al} \cite{4}.

The Hamiltonian for this system reads (\cite{11}, paragraph 9.3.2 and \cite{12}):
\begin{eqnarray}
H(R)=\left[\begin{array}{cc}
          \frac{\Delta\epsilon}{2}&\frac{C}{R^3}\\&\\
          \frac{C}{R^3}&-\frac{\Delta\epsilon}{2}\end{array}\right]
          \label{a}
\end{eqnarray}
where $\Delta\epsilon$ and $C$ are constants and $R=\sqrt{{b'}^2+v^2t^2}$ is the
distance between the atoms. Introducing
$d=(2C/\Delta\epsilon)^{\frac{1}{3}}$ as a natural distant unit for this case and
$T=d/v$ as the corresponding adiabatic
parameter and rescaling: $t\to sT$ and $b'\to bd$ we get from (\ref{a}):
\begin{eqnarray}
H(s)=\frac{\Delta\epsilon}{2}\left[\begin{array}{cc}1&
\frac{1}{(b^2+s^2)^{\frac{3}{2}}}\\&\\\frac{1}{(b^2+s^2)^{\frac{3}{2}}}&
-1\end{array}\right]
\label{b}
\end{eqnarray}

In the 'magnetic field' language we have of course ${\bf B}(sT,T)=\left(\left
(b^2+s^2\right)^{-\frac{3}{2}},0,1\right)\frac{\Delta\epsilon}{\mu}$ so that
all the assumptions ${\bf 1}^0-{\bf 8}^0$ above are satisfied with
${\bf B}^{\pm}(T)={\bf B}^{\pm}(\pm\infty,T)=(0,0,1)\frac{\Delta\epsilon}{\mu}$.

Obviously the last form of the ${\bf B}$-field shows that it belongs to the class of
Berman {\it et al} with two turning points on the "main" Stokes line (see Fig.1)
so that the formulae (\ref{232}) and (\ref{233}) are applicable readily.

\section{Discussion and conclusions}

\hskip+2em In our present calculations of the adiabatic limit for the transition
amplitudes in the two energy level systems we have corrected erroneous formulae of
 our previous paper \cite{01}. We have considered systems with the NED properties,
i.e. for which their corresponding Stokes graphs do not differ essentially from
their adiabatic
limit forms.

A formula (\ref{223}) which gives the corresponding transition amplitudes in the
adiabatic limit shows that these amplitudes result as an interference of
contributions coming from all complex conjugated pairs of turning points lying
on the same complex conjugated Stokes lines of the
respective limiting Stokes graph. Up to an overall phase it coincides with the one of
Joye, Mileti and Pfister \cite{4}.

A particularly simple formula for the transition amplitudes follows from a general
one (\ref{225}) when the latter is applied to the NED systems considered by Berman
{\it et al} \cite{20} with two turning points on the "main" Stokes line.
Namely, it obtains then the form (\ref{232}) found earlier
by Nikitin and Umanskii
\cite{11} as well as by Crothers \cite{16} and by Davies and Pechukas \cite{19} using
the steepest-descent methods.

\section*{Acknowledgments}

\hskip+2em This work has been supported by the KBN grant No. 5P03B06021


\begin{thebibliography}{99}
\bibitem{01} Giller S., Gonera C., {\it Phys. Rev.} {\bf A63} 052102 (2001)
\bibitem{1} Landau L.D., Lifshitz E.M., {\bf Quantum Mechanics. Nonrelativistic Theory.},
                                       (Pergamon: New York, 1965)
\bibitem{2} Davidovich L., {\it Quantum Optics in Cavities and the Classical Limit of Quantum Mechanics},
in: {\bf Latin-American School of Physics XXXI ELAF. New Perspectives
                                        on Quantum Mechanics 1998}
 Ed. Shahen Hacyan, Rocio Jauregui (AIP, Woodburg, New York 1999)

     Haroche S., {\it Cavity Quantum Electrodynamics: A Review of Rydberg Atom -
    Microwave Experiments on Entanglement and Decoherence}, ibidem
\bibitem{3} Joye A., Kunz H., Pfister Ch.-Ed., {\it Ann. Phys.} {\bf 208}, 299 (1991)
\bibitem{4} Joye A., Mileti G., Pfister Ch.-Ed., {\it Phys. Rev.} {\bf
    A44} 4280 (1991)
\bibitem{5} Joye A., Pfister Ch.-Ed., {\it J. Phys.} {\bf 24} 753 (1991)
\bibitem{6} Joye A., Pfister Ch.-Ed., {\it Phys. Lett.} {\bf A169} 62 (1992)
\bibitem{7}  Joye A., Pfister Ch.-Ed., {\it J. Math. Phys.} {\bf 34} 454 (1993)
\bibitem{8}  Fr\"oman N. and Fr\"oman P. O., {\bf JWKB Approximation. Contribution to the Theory},
                               North-Holland, Amsterdam 1965
\bibitem{9} Fedoriuk M. V., {\bf Asymptotic Methods for Linear Ordinary Differential Equations}
                                                 (Nauka: Moscow, 1983 (in Russian))
\bibitem{10} Giller S., {\it Acta Phys. Pol.} {\bf B23} 457 (1992),
{\it J. Phys.} {\bf A33} 1543 (2000)

     Giller S., Milczarski P., {\it J. Phys.} {\bf 32} 955 (1999), {\it J. Phys.} {\bf 33} 357 (2000)
\bibitem{13} Landau L.D., {\bf Collected Papers of L.D. Landau} (Pergamon, Oxford, 1965)
\bibitem{14} Zener C., {\it Proc. R. Soc.} {\bf 137} 696 (1932)
\bibitem{15} Dykhne A.M., {\it Zh. Eksp. Teor. Fiz.} {\bf 41}, 1324 (1961)
\bibitem{17} Lange S., {\bf Complex Analysis} (Springer, New York, 1993)
\bibitem{18} Langer R. E., {\it Phys. Rev.} {\bf 51} 669 (1937)
\bibitem{20} Berman P.R., Lixin Yan, Keng-Hwee Chiam, Ruwang Sung, {\it Phys. Rev.}
{\bf A 57} 79 (1998)
\bibitem{11} Nikitin E.E., Umanskii S. Ya., {\bf Theory of Slow Atomic Collisions} (Springer-Verlag, Berlin,                                                            1984)
\bibitem{12} Nikitin E.E., {\it Chem. Phys. Lett.} {\bf 2} 402 (1968)
\bibitem{16} Crothers D.S.F., {\it J. Phys.} {\bf A 5} 1680 (1972); {\it J. Phys.} {\bf B 6}
1418 (1973)
\bibitem{19} Davis J.P. and Pechukas P., {\it J. Chem. Phys.} {\bf 60} 3129 (1976)
\end{thebibliography}
\end{document}